\documentclass[%
onecolumn,
 amsmath,amssymb,
 aps,
prb,
]{revtex4-2}

\usepackage{graphics,graphicx} 
\usepackage{dcolumn} 
\usepackage{bm} 
\usepackage{xcolor}

\usepackage{amsmath,amssymb,amsfonts} 
\usepackage{textcomp}
\usepackage{color,soul} 
\usepackage{xfrac}
\usepackage{hyperref} 
\hypersetup{colorlinks=true, citecolor=blue, urlcolor=blue, linkcolor=blue}

\def\Iabs{\ensuremath{\langle I_\mathrm{abs} \rangle}} 
\def\Ieff{\ensuremath{I_\mathrm{eff}}} 
\def\AGeff{\ensuremath{A_\mathrm{eff}^G}} 
\def\Ec{\ensuremath\mathcal{E}} 
\def\Eimp{\ensuremath\Ec_\mathrm{imp}} 
\def\Epd{\ensuremath\Ec^{D}} 
\def\DOS{\ensuremath N_\mathrm{DOS}} 
\newcommand\Real[1]{\ensuremath{\mathrm{Re}\{#1\}}}
\newcommand\Imag[1]{\ensuremath{\mathrm{Im}\{#1\}}}
\newcommand\sig[1]{\ensuremath{\sigma^{(#1)}}}
\newcommand\fv[1]{\ensuremath{\tilde{\mathbf{#1}}}} 
\newcommand\tv[1]{\ensuremath{\mathbf{#1}}} 

\DeclareMathAlphabet\mathbfcal{OMS}{cmsy}{b}{n}

\begin{document}

\preprint{APS/123-QED}

\title{Graphene optical nonlinearity: From the third-order to the non-perturbative electrodynamic regime}

\author{Alexandros Pitilakis}
\email[Electronic mail: ]{alexpiti@auth.gr}
\affiliation{School of Electrical and Computer Engineering, Aristotle University of Thessaloniki}
\author{Emmanouil E. Kriezis}
\affiliation{School of Electrical and Computer Engineering, Aristotle University of Thessaloniki}

\date{\today}

\begin{abstract}
A non-perturbative model for graphene optical nonlinearity is developed for the study of ultrafast pulse propagation along a monolayer, as in the case of graphene-comprising nanophotonic integrated waveguides. This graphene `hot electron' model (GHEM) builds upon earlier work, based on the Fermi-Dirac framework for 2D semiconductors, which was aimed mainly at steady-state absorptive response of a monolayer under free-space laser-beam illumination. Our extension adapts the GHEM to in-plane light-matter interaction along graphene monolayers under intense ps-pulse excitation that leads to the carrier-density saturation regime. We first provide a quantitative overview of the `classic' perturbative third-order nonlinear response and then study the static and transient response of graphene as a function of the GHEM parameters, with focus on the monolayer quality and voltage-tunability. These results are compared to phenomenological models for saturable absorption and nonlinear refraction, showing good agreement with recent experimental works. In conclusion, the GHEM unifies the experimentally observed absorptive and refractive optical nonlinearities in a single multi-parametric framework therefore enabling the evaluation of the voltage-tunable light-matter interaction in structures such as diffraction-limited nanophotonic waveguides. This formalism can be readily employed to THz frequencies, adapted to multi-channel (e.g. pump-probe) nonlinear effects, or developed to include carrier and lattice-temperature diffusion to extend its validity threshold.
\end{abstract}

\maketitle

\tableofcontents

\section{Introduction} \label{sec:1:intro}
Graphene monolayers \cite{Ferrari2015} exhibit a broadband nonlinear response \cite{Cheng2018}, from the MIR/THz to the NIR/VIS spectral region, with diversified absorptive and refractive features, which are moreover controllable thanks to the material's high electro-optic tunability \cite{Cox2014}. More than a decade ago intense research on the subject ignited mostly due to experimental evidence in \cite{Hendry2010}, which corroborated the first theoretical predictions in \cite{Mikhailov2007,Mikhailov2008,Ishikawa2010}. Various turns of events have marked the way, including orders-of-magnitude shifts in the absolute values reported, sign changes, and wide deviations in underlying model parameters. Despite this turbulent course, all high-power evidence \cite{Zhang2012,Hong2013,Yang2018,Lin2018} indicates that graphene possesses a large magnitude, broadband and tunable nonlinearity, with plenty of untapped potential both in theory and in practical implementation. Naturally, interest shifted from free-space optics to nanophotonics \cite{GarcadeAbajo2013} and integrated photonics \cite{Driscoll2012,Driscoll2015}, a platform where light-graphene interaction can be maximized, owing to the diffraction-limited confinement that guided modes can routinely provide \cite{Pitilakis2013,Pitilakis2016}. Special interest is given to on-chip pulsed laser platforms \cite{Yao2018OnChipGrapheneComb,Mock2017}, where cavity-buildup can be elevated to its extreme and the interplay between refractive and absorptive nonlinearities is ever more crucial to the system response.

Nonlinear response of 2D semiconductors like graphene can be distinguished in two regimes, firstly the perturbative (or parametric), where nonlinear polarization quasi instantaneously follows the applied field, and secondly the non-perturbative (or electrodynamic) regime, which is related to free-carrier effects as in bulk semiconductors. The carrier-related effects in graphene have been extensively studied in focused laser beam experiments \cite{Breusing2011,Winnerl2011,Malic2011,Cox2017,Baudisch2018,Hafez2018,Hafez2019}, where NIR illumination is mostly considered as a pump mechanism, exciting nonlinear effects which are probed by THz pulse. Back to the perturbative regime, more relevant to integrated photonics, complicated formulas have been theoretically extracted for the associated third-order surface conductivity in \cite{Cheng2014,Cheng2015,Mikhailov2016Quantum}, considering monolayers in free-space, and experiments \cite{Dremetsika2016,Jiang2018} have validated and slightly twisted the predictions, especially regarding the sign of the refractive nonlinearities. Theoretical studies \cite{Chatzidimitriou2015,Pitilakis2016,Ooi2014} and experiments \cite{Vermeulen2016,Alexander2017,Alexander2018,Vermeulen2018,Demongodin2019} in integrated nonlinear waveguide devices followed, which paved the way toward voltage-tunable ultrafast absorptive and refractive nonlinearity in photonics. While the interested audience awaits for more integrated nonlinear devices and experiments, work progresses in designing new components  \cite{Wang2015SiNDetector,AbdollahiShiramin2017ModSwitch,Chatzidimitriou2018,Doukas2018,Ono2019,Chatzidimitriou2020,ZhangMoss2021,Pitilakis2021,Chatzidimitriou2021,Sahoo2021} and extending the theoretical modeling horizons \cite{Semnani2016,Semnani2017,Marini2017,Semnani2019,Mikhailov2019HEM,Soavi2019,Cheng2020,Mikhailov2021}.

\begin{figure}[h]
    \centering
    \includegraphics{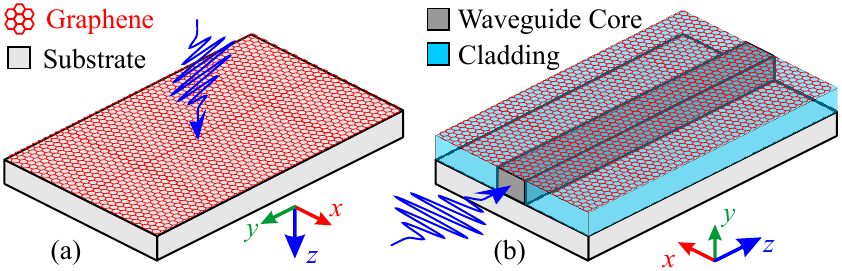}
    \caption{High-power pulse (a) normally impinging on an infinite graphene monolayer lying on a thick dielectric substrate, (b) propagating along an integrated nanophotonic waveguide whose mode spatially overlaps with a graphene monolayer.}
    \label{fig:1:Schems}    
\end{figure}

This work presents a comprehensive overview and a comparative assessment of graphene nonlinearity across diverse operation regimes, and elucidates issues related to the exploitation of the Kerr-effect and saturable absorption along integrated NIR photonic waveguides. For this reason, we mainly focus on the effect of the voltage-tunable chemical potential and the fabrication quality of graphene monolayers. We first briefly visit the classic linear and third-order nonlinear regimes, collectively referred to as perturbative response. We then move to the strongly nonlinear non-perturbative regime within the framework of an electrodynamic model obeying Fermi-Dirac statistics which we refer to as \textit{graphene `hot electron' model} (GHEM). This model is based on foundational work in \cite{Mikhailov2019HEM}, which was devised for free-space laser-beam transmission through a graphene monolayer, Fig.~\ref{fig:1:Schems}(a), mainly studying absorption change in the steady-state/static (continuous wave, CW) excitation. In this work, we adapt and modify the GHEM for transient analysis, i.e., for quantifying the nonlinear response to picosecond pulses propagating along graphene-comprising waveguides, Fig.~\ref{fig:1:Schems}(b). Moreover, we consider intensities corresponding to the photogenerated carrier-density saturation regime and focus on the coupled refractive and absorptive nonlinear effects. Our results are in good qualitative agreement with recent experimental results, in both free-space and guided-wave configurations, and provide a range of parameters that can be fit to measurements for quantitative evaluation. This GHEM extends the work of \cite{Mikhailov2019HEM}, to the study of ultrafast pulse propagation in highly nonlinear integrated waveguides, providing physical insight to the analysis and design guidelines. Extension of this formalism to other nonlinear 2D materials \cite{Ferrari2015,YouPanoiu2018NL2D,Cheng2018} can also be envisaged.

The rest of the paper is organized as follows: In Section~\ref{sec:2:Classical} we set out from the classical linear and perturbative third-order nonlinear regime, highlighting their pertinent aspects for pulse propagation in graphene-comprising waveguides beyond quasi-equilibrium. Section~\ref{sec:3:GHEMtheory} briefly outlines the GHEM model theoretical framework and \ref{sec:4:GHEMresults} presents a comprehensive study of the parametrically evaluated static (CW) and transient (pulsed) response. Section~\ref{sec:5:Discussion} discusses our results' implications for photonic waveguides, assesses qualitative trends, compares to recent experiments, and outline possible extensions of the model. Finally, following the concluding remarks of Section~\ref{sec:6:Summary}, a set of Appendices contain the bulk of the model formulas and elaboration on their less intuitive aspects, together with auxiliary proofs and numerical implementation details.

\section{Linear and Perturbative Third-Order Nonlinear Regime} \label{sec:2:Classical}
Graphene's EM response to wavelengths below UV is well described by surface conductivity tensors $\tilde{\sigma}^{(n)}$, denoting a 2D/sheet material, rather than with susceptibility tensors $\tilde{\chi}^{(n)}$, which characterise 3D/bulk materials; this is due to the 0.35~nm thickness of the graphene monolayer. 

In the absence of strong magnetic biasing, graphene's surface conductivity is isotropic and, thus, it can be fully characterized by a single scalar complex value, $\sig{n}$. As a 2D/sheet medium, graphene only interacts with electric fields tangential to their surface, giving rise to a surface current density according to Ohm's law, $\tv{J}_s=\sig{1}_T\tv{E}_\parallel$, in units of A/m. In this expression, $\sig{1}_T=\sig{1}_\mathrm{lin}+\sig{1}_\mathrm{NL}$ is the \textit{total} surface conductivity including both linear and nonlinear contributions. The nonlinear contribution depends on the local tangential E-field magnitude, $|\tv{E}_\parallel|$, and other material-depended parameters, and can fall in perturbative and non-perturbative regimes; the former is the subject of this section while the latter is addressed by the nonlinear GHEM developed in the following sections.

\subsection{Light-Graphene Interaction}

When interacting with EM waves of small photon energies, graphene behaves like a {zero-bandgap 2D semiconductor} with a linear energy-momentum band structure, as opposed to bulk/3D semiconductors which have nonzero bandgap and parabolic band structure. As a semiconductor, graphene's electro-optical response can be deduced from Fermi-Dirac statistics, governed by the Pauli exclusion principle, in terms of a common `hot' carrier temperature ($T$) and the quasi Fermi-levels for the conduction ($\mu_e$) and valence ($\mu_h$) band occupation, corresponding to electron and hole plasmas. Under thermal equilibrium, i.e., in the absence of strong optical illumination, it holds that $T\equiv T_0$, where $T_0$ is the lattice temperature, and $\mu_e\equiv\mu_h=\mu_c$, where $\mu_c$ is the chemical potential of graphene. At zero temperature $\mu_c=\Ec_F$, i.e., the chemical potential is equal to the Fermi energy of graphene (where the Fermi-Dirac carrier distribution function equals 0.5), whereas at room temperature it is only slightly smaller; consequently, quantities $\mu_c$ and $\Ec_F$ are sometimes used interchangeably. Negative or positive $\mu_c$ denotes the holes or the electrons as majority carriers, respectively. However, the two types of carriers are symmetric in graphene (`ambipolar' electron and hole transport), meaning that $\mu_c=\pm|\mu_0|$ will theoretically have identical optical properties, which is the case throughout this work. Nevertheless, graphene fabrication techniques (e.g., CVD growing or exfoliation) and its environment (e.g., attachment to dielectric or semiconductor substrates) lead to either positive or negative charge densities. For more details on the Fermi-Dirac framework, refer to Appendix~\ref{app:FD_Theory}.

\begin{figure}[h]
    \centering
    \includegraphics{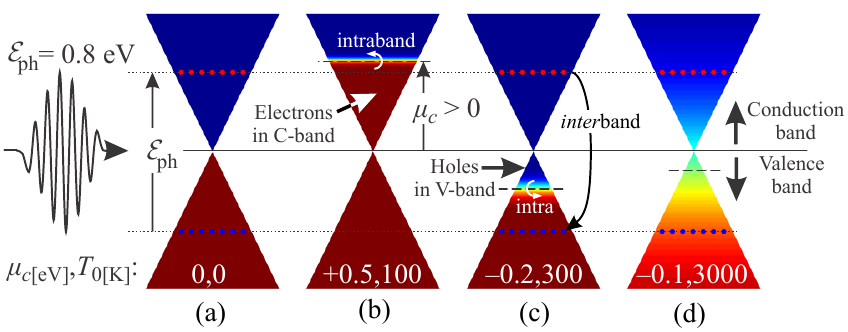}
    \caption{Energy-momentum graphene cones depicting band-filling at equilibrium lattice temperature $T_0$. The colormap denotes the distribution function of carriers in the two bands and the horizontal dashed lines denote the chemical potential, $\mu_c$. (a) Pristine graphene at zero temperature. (b) High electron density at 100~K. (c) Low hole density at 300~K. (d) Quasi-unbiased graphene super heated at 3000~K. Interaction with 0.8~eV photons gives rise to both intraband and interband surface conductivity in all cases, except case (b), where Pauli blocking cancels the interband mechanism.}
    \label{fig:2:cones}    
\end{figure}

Graphene electro-optical properties can be abstracted in Dirac cone diagrams such as the ones in Fig.~\ref{fig:2:cones}, depicting band filling, Eq.~\eqref{eq:feh}, at thermal equilibrium and for a few characteristic $\{T,\mu_c\}$ combinations. It can be inferred that graphene is a semi-metal, i.e., it can act either as a conductor or a dielectric, if its $|\mu_c|$ is below or above the half-photon energy, $\hbar\omega/2$, respectively. In its dielectric regime ($|\mu_c|\gg\hbar\omega/2$), graphene is highly transparent in the NIR and above. But, in its metallic regime ($|\mu_c|\ll\hbar\omega/2$), it can exhibit remarkably high conductivity for its sub-nm thickness, from the THz up to the visible, and even plasmonic behaviour (i.e., equivalent permittivity with $\Real{\varepsilon_\mathrm{r,eq.}}\ll-1$) in the THz. Evidently, control over graphene's $\mu_c$ (or equivalently its carrier density, Appendix~\ref{app:FD_Theory}) exerts control over its optical response, i.e., over the surface conductivity that an electromagnetic (EM) field experiences when in proximity to the material. There are various schemes for tuning $\mu_c$, either at fabrication or electrically, with biasing or gating, using appropriate electrode configurations \cite{Alexander2018}.

Figure~\ref{fig:2:cones} additionally provides the groundwork for the nonlinear GHEM, i.e., depicts the two mechanisms with which graphene interacts with optical photons: Intraband absorption shifts carriers away from the Dirac point within the same band whereas interband absorption of photons with $\Ec_\mathrm{ph}=\hbar\omega > 2\mu_c$ (above Pauli blocking) moves carriers between bands, depicted as small colored dots in Fig.~\ref{fig:2:cones}. Evidently, the intraband mechanism is equivalent to a carrier temperature shift (heating and cooling) whereas the interband mechanism is responsible for the photogeneration of electron-hole pairs (satisfying electro-neutrality condition, in the absence on carrier injection) which eventually recombine. At low-power excitation, these two mechanisms act \textit{perturbatively} meaning that the graphene's quantum state, i.e., its $\{T,\mu\}$ properties, remain unchanged (thermal equilibrium is instantaneously reinstated) and the response is assumed instantaneous and fully described by a limited set of $\sig{n}$ conductivities in a Taylor-like expansion \cite{ButcherCotter1990,Boyd2020}. In contrast, at sufficiently high excitation, graphene's quantum state is changed, i.e., its $\{T,\mu\}$ properties are \textit{non-perturbatively} affected, which leads to a nonlinear response, strongly related to the free-carrier (plasma) spatiotemporal dynamics.

In this work, we are concerned primarily with graphene's behaviour in the NIR, $\hbar\omega\approx0.8$~eV ($\lambda_0\approx1550$~nm), and for $|\mu_c|$ close to and below the half-photon energy; this is interesting for three reasons: (i) due to the switching of graphene's \textit{qualitative} response manifested around that threshold, (ii) because of the non-trivial physics required to \textit{quantitatively} evaluate its nonlinear properties, and (iii) as most state-of-the-art fabricated graphene-comprising waveguide devices for the NIR have $|\mu_c|$ in that vicinity, when unbiased, and a few-Volt potential can shift it to cover the whole meaningful range $\mu_c=\pm\hbar\omega$.

\subsection{Linear Response}

The linear response of graphene to a low intensity EM wave at harmonic frequency $\omega_0$ can be extracted by a Kubo-like semi-classical formalism, i.e., tight-binding in the single-atom approximation. There are multiple pathways that lead to the same formulas \cite{Falkovsky2008,Cheng2014,Cheng2015,Mikhailov2016Quantum}, which produce the total linear surface conductivity of graphene, $\sig{1}=\sig{1}_i+\sig{1}_e$, as a function of parameters $\{\omega_0,\mu_e,\mu_h,T\}$ and the momentum relaxation rates $\Gamma_{i,e}=\hbar/\tau_{i,e}$. Subscripts $\{i,e\}$ denote $\{$intraband, interband$\}$ throughout the document; the rates ($\Gamma$) are usually expressed in meV units and the corresponding lifetimes ($\tau$) in fs, where the $\tau_\mathrm{[fs]}\Gamma_\mathrm{[meV]}\approx 658$ relation is handy for conversion. 

At finite (nonzero) temperature, the formulas for the $i$- and $e$-conductivities entail integration over the energy spectrum, accounting also for possible energy dependence of the momentum relaxation rates, $\Gamma_{i,e}=\Gamma_{i,e} (\Ec)$. This is required to model and balance the effect of various scattering mechanisms that mediate the carrier relaxation, namely scattering of carriers from impurities, phonons, and other lattice imperfections. This dependence is of particular importance for the intraband term and for high temperatures, where most experiments evidence a dominance of `long-range' scattering with charged impurities (rather than hot phonons) \cite{Hafez2019}; specifically, the relaxation lifetime is proportional to the energy when $\Ec>\Eimp$, where $\Eimp$ is a Coulomb energy that increases with the square root of the density of impurities \cite{Mikhailov2019HEM}. More details, together with the full and simplified expressions for the numerical computation of $\sig{1}_{i,e}(\omega_0,\mu_e,\mu_h,T,\Gamma_{i,e,})$, can be found in the Appendix~\ref{app:sigmaCalc}. Note that $\Real{\sig{1}}>0$ holds for absorptive part of the total surface conductivity, whereas the refractive part, $\Imag{\sig{1}}$, can be positive or negative. In the NIR, the effect of graphene sheets is mostly perturbative, especially for the refractive part; however, in the THz band, $\Imag{\sig{1}}\gg1$ which gives rise to a plasmonic behaviour, $\Real{\varepsilon_\mathrm{r,eq.}}\ll-1$.

\begin{figure}[h]
    \centering
    \includegraphics[width=80mm]{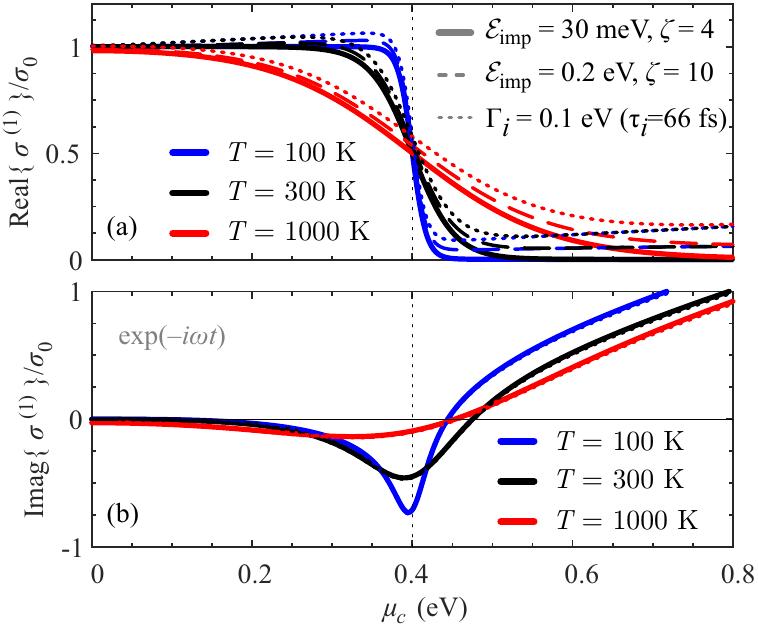}
    \caption{(a) Real and (b) imaginary part of the normalized total $\sig{1}/\sigma_0$ vs. $\mu_c$ at $\lambda_0=1550$~nm (0.8~eV). The interband momentum relaxation rate is fixed at $\Gamma_e=0.5$~meV ($\tau_e=1.3$~ps) and the intraband $\Gamma_i$ is either constant (dotted line) or energy dependent according to the model of Eq.~\eqref{eq:Gintra}.}
    \label{fig:s1Linear}    
\end{figure}

Assuming thermal equilibrium, $\mu_e=\mu_h=\mu_c$ and $T=T_0$, and using Eq.~\eqref{eq:s1i_full} for the intraband and \eqref{eq:s1e_integrable} for the interband conductivity, we present in Fig.~\ref{fig:s1Linear} a brief parametric investigation of the total surface conductivity against the parameters pertinent to this study: chemical potential ($\mu_c$), temperature ($T$), and intraband momentum relaxation rate ($\Gamma_i$). $\sig{1}$ is normalized to the universal surface conductivity, $\sigma_0=q^2/4\hbar\approx61~\mu$S, where $q$ is the electron charge and $\hbar$ the reduced Plank constant. We note the high absorption when $\mu_c<\hbar\omega/2=0.4$~eV (interband absorption allowed), the smoothing of the spectral features as the carrier temperature increases, and the residual absorption slope for $\mu_c>0.4$~eV which gets larger as the graphene quality drops ($\Gamma$ increases). This residual absorption is caused by the intraband mechanism at high carrier densities and is quantified either by a high (constant) $\Gamma_i$ or by a high impurity energy $\Eimp$; more details can be found in the Appendix subsection~\ref{sec:s1intra}. Now, concerning the imaginary (refractive) part of $\sig{1}$, we note that it is only affected by temperature near the half-photon energy, it exhibits a sign-flip coinciding with the absorption drop, and it scales linearly with $|\mu_c|$ implying a proportionality to $\sqrt{n_{T}}$, where $n_{T}=n_e+n_h$ is the total carrier density, Eq.\eqref{eq:neh}. Finally, we note that the curves of Fig.~\ref{fig:s1Linear} are even-symmetric around the Dirac point, $\mu_c=0$, and that sometimes the horizontal axis is normalized, i.e., $(\mu_c/\hbar\omega)^{\pm1}$; in this sense the term `$\mu_c$ spectra' is used in this work as $\omega$ is fixed.

The useful conclusions from this analysis are: (i) that good quality graphene, such as $\Eimp=30$~meV corresponding to 7000~cm$^2$/Vs mobility, can be switched between the opaque and transparent regime by tuning $|\mu_c|$ from $0.3$ to $~0.5$~eV at room temperature, and (ii) that the refractive part of the conductivity scales with the square-root of carrier density and is virtually unaffected by sample quality.

\subsection{Perturbative Third-Order Response}

The first attempts to theoretically evaluate the nonlinear response of graphene naturally focused on its third-order surface conductivity, $\sig{3}$, because in the absence of strong magnetic biasing it holds that $\sig{2}=0$, owing to graphene's centrosymmetric crystal lattice \cite{ButcherCotter1990}. Moreover, the generalized fourth-rank tensor $\tilde{\sigma}^{(3)}_{abcd}(\omega_0;\omega_1,\omega_2,\omega_3)$, with $\{a,b,c,d\}=\{x,y,z\}$ and $\omega_0=\omega_1+\omega_2+\omega_3$, is isotropic, meaning that only one of its elements, $\sig{3}_{xxxx}=\sigma_3$ by convention, suffices for population of the full tensor.

It must be noted that $\sigma_3$ is complex: the sign of its real part corresponds to induced absorption ($+$) or induced transparency/absorption saturation ($-$), while the sign of the imaginary part corresponds to focusing ($-$) or defocusing ($+$) refractive Kerr-type nonlinearity, under the $e^{-i\omega t}$ phase convention. Here, we must stress that a positive or negative $\Imag{\sigma_3}$ only \textit{approximately} corresponds to defocusing or focusing nonlinear refraction, respectively. The proportionality holds only for large $|\Imag{\sigma_3}|$ and low-loss media (low $\Real{\sig{1}}$ in the case of graphene), while the exact value of the real part of the \textit{equivalent} nonlinear index ($\Real{n_2}$) and its sign-transition is governed by more complicated expressions that also account for complex-valued $\sig{1}$, similar to the case of bulk nonlinear media with losses \cite{delCoso2004}. The same approximation holds for the sign of $\Real{\sigma_3}$, roughly corresponding to the sign of $\Imag{n_2}$.

Early estimates for $\sigma_3$ were triggered by remarkably high experimentally measured four-wave mixing \cite{Hendry2010}, followed by a set of thorough quasi-classical single-atom derivations (perturbative solution of semiconductor Bloch equations in 2D), starting from zero temperature and zero relaxation rate \cite{Cheng2014} and later moving to finite temperature and relaxation with distinct $\Gamma_i\neq\Gamma_e$ \cite{Cheng2015}. These results were immediately corroborated by a different quantum electrodynamic model \cite{Mikhailov2016Quantum}, applicable only for $\Gamma_i\equiv\Gamma_e$ but avoiding the divergent behaviour as $\mu_c\rightarrow 0$; this was initially a zero-temperature model that was later extended to finite temperature \cite{Savostianova2018} with a special focus on the Kerr/two-photon absorption (TPA) effect. For identical inputs, $\{\mu_c,T,\Gamma_i\equiv\Gamma_e$\}, the two models produce equivalent results for $\Real{\sigma_3}$ but their convergence in the refractive part, $\Imag{\sigma_3}$, is limited to high $\Gamma>20$~meV or above one-photon energy, $\mu_c>\hbar\omega$. In any case, both models can be cast to the study of various third-order phenomena, such as Kerr/TPA, third-harmonic generation ($\omega_0/3=\omega_1=\omega_2=\omega_3$), or parametric frequency conversion ($\omega_1\neq\omega_2=\omega_3$ so that an idler at $\omega_0=2\omega_\mathrm{pump}-\omega_\mathrm{sig}$ is generated by a weak and strong signal at different wavelengths). It should be noted that despite the isotropic nature of the $\tilde{\sigma}^{(3)}$ tensor, extracting the complex scalar $\sigma_3=\sigma_3(\omega_0,\mu_c,T,\Gamma)$ in its customary symmetrized form requires complicated calculations of multiple unsymmetrized tensor components \cite{Cheng2015,Mikhailov2016Quantum}.

\begin{figure}[h]
    \centering
    \includegraphics[width=80mm]{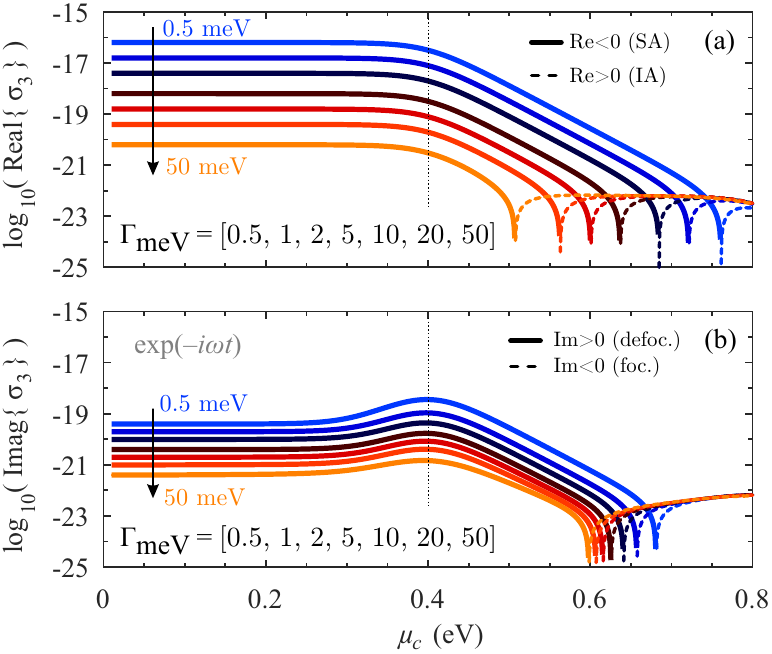}
    \caption{Effect of relaxation rate $\Gamma_i=\Gamma_e$ on $\mu_c$ spectra, at $\lambda_0=1550$~nm (0.8~eV) and $T=300$~K. (a) Real part of $\sigma_3$, with its sign denoting saturable absorption (SA) or induced absorption (IA). (b) Imaginary part of $\sigma_3$, with its sign denoting defocusing or focusing refractive nonlinearity.}
    \label{fig:s3vsmu_varGi}    
\end{figure}

\begin{figure}[h]
    \centering
    \includegraphics[width=80mm]{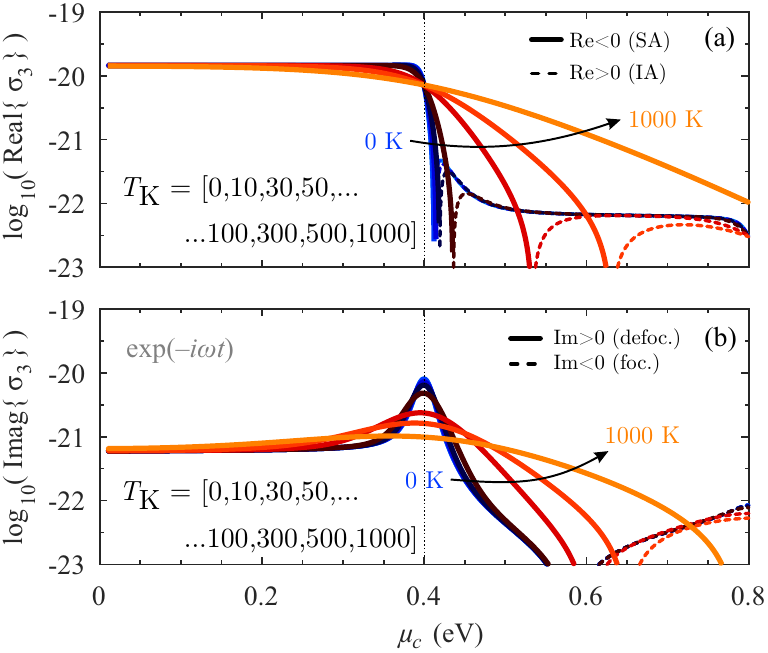}
    \caption{Effect of lattice temperature on $\mu_c$ spectra at $\lambda_0=1550$~nm (0.8~eV) and $\Gamma_i=\Gamma_e=33$~meV, i.e., $\tau=20$~fs. (a) Real part of $\sigma_3$, with its sign denoting saturable absorption (SA) or induced absorption (IA). (b) Imaginary part of $\sigma_3$, with its sign denoting defocusing or focusing refractive nonlinearity.}
    \label{fig:s3vsmu_varTemp}    
\end{figure}

In this work, we are interested in single-channel self-acting nonlinear phenomena, so we will solely study nonlinear conductivity related to the Kerr effect, $\sigma^{(3)}_{xxxx}(\omega;-\omega,+\omega,+\omega)$, and evaluate its dependence on graphene parameters $\{\mu_c,T,\Gamma\}$ at $\lambda_0=1550$~nm (0.8~eV). Using the model of \cite{Mikhailov2016Quantum} we study the $\mu_c$ spectra of $\sigma_3$, equal to $\sigma^{(3)}_\mathrm{Kerr}$ normalized with respect to 1~S(m/V)$^2$, for a few pertinent parameter combinations; note that, similarly to $\sig{1}$, the $\sig{3}$ spectra are also even-symmetric for negative $\mu_c$. In Fig.~\ref{fig:s3vsmu_varGi} we study the effect of chemical potential and scattering rates $\Gamma_i=\Gamma_e$, at room temperature (300~K): For $\mu_c<\hbar\omega/2$, in all cases we get SA and defocusing refraction (DFR), both `flat' and vanishing as $\Gamma_{i,e}$ increase, i.e., as the quality of the graphene sample decreases. Above half-photon energy, the magnitude of nonlinearity decreases and experiences sign changes.
In Fig.~\ref{fig:s3vsmu_varTemp} we study the effect of chemical potential and temperature, for a typical value of $\Gamma_i=33$~meV ($\tau_i=20$~fs). We notice that the increasing temperature effect is almost exclusively near and above half-photon energy, extending the SA and DFR regime to higher chemical potentials; note the peaking of $\Imag{\sigma_3}$ on half-photon energy, whose resonance increases as temperature decreases.

Summarizing this subsection, for the typical case of room temperature and $\mu_c\approx0.2$-$0.3$~eV, one expects SA and DFR, from the real and imaginary parts of $\sigma_3$, respectively, with low sensitivity on $\mu_c$. Moreover, the magnitude of perturbative nonlinearity depends strongly on the momentum relaxation lifetimes: As the sample quality improves ($\Gamma_{i,e}\rightarrow0$) both SA and DFR increase by orders or magnitude. This is the reason for the logarithmic scaling used in the vertical axes of Fig.~\ref{fig:s3vsmu_varGi} and \ref{fig:s3vsmu_varTemp}.

\section{Nonperturbative Regime: Theory} \label{sec:3:GHEMtheory}
The theory and governing equations of the GHEM will be outlined in this section. We set off from the comprehensive model developed by S.~A.~Mikhailov in \cite{Mikhailov2019HEM}, and adapt it to the study of ultrafast pulse propagation in graphene-comprising nonlinear waveguides. In all this analysis, we have adopted the Fermi-Dirac framework described in Appendix~\ref{app:FD_Theory} for modeling the carrier plasma in graphene under intense optical excitation that pushes the medium out of thermal equilibrium.

\subsection{Mikhailov's Hot Electron Model}

The GHEM developed in \cite{Mikhailov2019HEM} relies firstly on instantaneous absorption, i.e., as if the carrier-carrier scattering time is zero, and secondly on the splitting of the effects arising from the intraband and interband absorption contributions. This is based on the assumption that the timescales of the two effects are different so that they can be assumed to act independently on the carrier plasma. Specifically, the intraband absorption, proportional to $\Real{\sig{1}_i}$ from Eq.~\eqref{eq:s1i_full}, instantaneously heats the carrier plasma which subsequently cools down in a $\tau_\Ec$ timescale, typically sub-ps and higher than the intraband momentum relaxation lifetime, $\tau_i=\hbar/\Gamma_i$. The interband absorption, proportional to $\Real{\sig{1}_e}$ from Eq.~\eqref{eq:s1e_full}, instantaneously generates electron-hole pairs which subsequently recombine in a longer few-ps timescale, i.e., $\tau_\mathrm{rec}\gg\tau_\Ec$; note that $\tau_\mathrm{rec}$ has the meaning of recombination time only at very low excitation intensities. In \cite{Mikhailov2019HEM}, it was found that the ratio between these two phenomenological lifetimes governs the response; in this work we set $\tau_\mathrm{rec}=10\tau_\Ec$, unless otherwise stated. The two absorption mechanisms can thereby be considered to act separately on graphene, and form two rate equations: one for the total energy density in the hot plasma, $\Epd_T$, and one for the photogenerated carrier density, $n_\mathrm{PG}$. Finally, as a transitional state between these effects, the GHEM assumes that a \textit{quasi}equilibrium (QE) state is instated shortly after illumination, where: (i) the plasma is characterised by QE chemical potentials, $\mu_{e_0}\neq\mu_{h_0}$, which are moreover different from graphene's effective hot $\mu_{e,h}$, (ii) the photogenerated carriers have not recombined, and (iii) the plasma has cooled down to the lattice temperature, $T\approx T_0$.

The GHEM in \cite{Mikhailov2019HEM} has six unknowns: $\{\Epd_T, n_\mathrm{PG},\mu_e, \mu_h, \mu_{e_0}, \mu_{h_0} \}$, and so requires six equations. The first two are rate equations related to the aforementioned intraband and interband absorption, as
\begin{equation}
    \frac{\partial\Epd_T}{\partial t} = \Iabs_i - \frac{ \Epd_T - \Epd_\mathrm{QE} }{\tau_\Ec},
    \label{eq:PDE_SAM_Energy}
\end{equation}
\begin{equation}
    \frac{\partial n_\mathrm{PG}}{\partial t} = \frac{\Iabs_e}{\hbar\omega} - \frac{n_\mathrm{PG}}{\tau_\mathrm{rec}} 
    \left(1 + \frac{n_\mathrm{PG}}{n_{T_0}}\right),
    \label{eq:PDE_SAM_nPG}
\end{equation}
respectively. In Eq.~\eqref{eq:PDE_SAM_Energy}, $\Epd_T=\Epd_e(\mu_e,T)+\Epd_h(\mu_h,T)$ is the total plasma energy density in the \textit{out-of-equilibrium} `hot' state, $\Epd_\mathrm{QE}=\Epd_\mathrm{QE}(\mu_{e_0},\mu_{h_0},T_0)$ is the corresponding energy density at the QE state, and $\tau_\Ec\leq1$~ps is the phenomenological energy-relaxation lifetime. In Eq.~\eqref{eq:PDE_SAM_nPG}, $n_{T_0}=n_e^0 + n_h^0 = n_e(\mu_c,\mu_c,T_0)+n_h(\mu_c,\mu_c,T_0)$ is the total carrier density at thermal equilibrium in the absence of illumination (where $\mu_e\equiv\mu_h=\mu_c$, the sample's chemical potential, and $T=T_0$, the lattice temperature), and $\tau_\mathrm{rec}\geq 10\tau_\Ec$ is the effective recombination (or, more abstractly, decay-rate) lifetime for the photogenerated carriers. The second term, $n_\mathrm{PG}/n_{T_0}$, in the parentheses in the RHS of Eq.~\eqref{eq:PDE_SAM_nPG} arises from the assumption that recombination rate is proportional to the difference of the \textit{product} of carrier densities between the hot and the QE state, $n_e n_h-n_e^0 n_h^0$, typical in semiconductors. Calculation formulas for the energy and carrier densities can be found in the Appendix~\ref{app:FD_Theory}.

Evidently, the evolution of graphene's photoconductive response depends on an effective absorbed intensity, $\Iabs_{i,e}$, measured in W/m$^2$. In \cite{Mikhailov2019HEM}, where we have the zero-dimensional assumption, i.e., a plane wave of sufficiently wide and uniform focal-spot as in Fig.~\ref{fig:1:Schems}(a), we simply define
\begin{equation}
    \Iabs_{i,e}=A_{i,e}I_\mathrm{inc}.
    \label{eq:Iabs_SAM_Tinkham}
\end{equation}
In this spectral domain expression, $I_\mathrm{inc}$ is the incident plane-wave intensity (in W/m$^2$) and $A_{i,e}$ is a (unitless) power absorption coefficient, that can be deduced from the `Tinkham formula' \cite{Tinkham1957}, for the perturbative case of an ultrathin conductive sheet in the interface between air and a semi-infinite dielectric substrate, as follows:
\begin{equation}
    A_{i,e}=A_{i,e}(\omega_0,\mu_e,\mu_h,T)=\frac{ 4 Z_0 \Real{\sig{1}_{i,e}}} {|n_\mathrm{sub}+1+\sig{1}Z_0|^2},
    \label{eq:TinkhamAbsorption}
\end{equation}
where $n_\mathrm{sub}$ is the substrate refractive index at harmonic frequency $\omega_0$ and $Z_0\approx 377$~Ohm is the free-space impedance. Note the full (complex) graphene conductivity, $\sig{1}=\sig{1}_i+\sig{1}_e$, in the denominator, as contrasted to the real part of either intra- or interband term in the nominator. These surface conductivity spectra depend on the hot (out-of-equilibrium) graphene plasma parameters, $\sig{1}_{i,e}=\sig{1}_{i,e}(\omega_0,\mu_e,\mu_h,T)$, with the formulas of Appendix~\ref{app:sigmaCalc}. The bandwidth of the nonlinear response in this model depends solely on $\sig{1}$, whose response is rather broadband as evidenced by Fig.~\ref{fig:s1Linear}, especially as temperature increases.

The remaining four equations of the GHEM, that complement the two rate equations, are algebraic and stem from electroneutrality (the number of photogenerated holes is equal to that of photogenerated electrons) in the hot and QE states:
\begin{subequations}\label{eq:mus_Electroneutrality}
\begin{align}
        \mu_e    &= +(k_B T) F_{1}^{-1}\left[\frac{\pi(\hbar v_F)^2}{( k_B T)^2} (n_e^0+n_\mathrm{PG})\right], \\
        \mu_h    &= -(k_B T) F_{1}^{-1}\left[\frac{\pi(\hbar v_F)^2}{( k_B T)^2} (n_h^0+n_\mathrm{PG})\right], \\
        \mu_{e_0} &= +(k_B T_0) F_{1}^{-1}\left[\frac{\pi(\hbar v_F)^2}{( k_B T_0)^2} (n_e^0+n_\mathrm{PG})\right], \\
        \mu_{h_0} &= -(k_B T_0) F_{1}^{-1}\left[\frac{\pi(\hbar v_F)^2}{( k_B T_0)^2} (n_h^0+n_\mathrm{PG})\right],    
    \end{align}
\end{subequations}
where $F_{1}^{-1}$ is the inverse of the Fermi-Dirac integral of first order, defined in Eq.~\eqref{eq:mu_from_n_T}.

For the numerical solution of the GHEM, in static (CW) or transient (pulsed) mode, refer to Appendix~\ref{app:Numerical_Tips}.

\subsection{Extensions and Modifications}

The contribution in this work includes three modifications to the GHEM of \cite{Mikhailov2019HEM}:

\subsubsection{Effective Absorbed Intensity} \label{sec:Iabseff}

For the application of the model on pulse \textit{propagation along} a graphene sheet, as in Fig.~\ref{fig:1:Schems}(b), and not \textit{impinging on} a sheet, as in Fig.~\ref{fig:1:Schems}(a), we first need to modify the effective absorbed intensity, $\Iabs$. In nanophotonic waveguides the incident intensity does not have a large and smooth spot, as in focused laser beams, but it has a mode distribution across the graphene sheet, close to the diffraction limit. In this case, we first define a \textit{local} absorbed intensity for narrowband quasi-harmonic fields \cite{Alexander2018,Chatzidimitriou2020}, 
\begin{equation}
    I_{\mathrm{abs},i/e}(\tv{r};t)=\frac{1}{2}\mathrm{Re}\{\sig{1}_{i/e}(\tv{r};t)\}|\tv{E}_\parallel(\tv{r};t)|^2,
    \label{eq:Iabslocal}
\end{equation}
where $t$ is a retarded timeframe (not the optical cycle), e.g., the pulse-envelope modulating a NIR carrier frequency $\omega$. Equation~\eqref{eq:Iabslocal} can be derived for time-harmonic versions of Ohm's law, $\fv{J}=\sigma_\mathrm{bulk}\fv{E}$, and the power-density resulting from Joule-heating in a volume $V$, $\partial P_J/\partial V = \frac{1}{2}\fv{E}\cdot\fv{J}^*$, adapted to sheet conductivity \cite{Chatzidimitriou2015,Chatzidimitriou2020}. From Eq.~\eqref{eq:Iabslocal}, we can subsequently abstract the on-chip pulse power and waveguide geometry, through the mode profile $\fv{e}(x,y)$ in the cross-section, to produce the effective $\Iabs_{i,e}(t)$ or $\Iabs_{i,e}(z,t)$ required when applying the pulsed GHEM to a free-space sheet or through a $z$-segment of a waveguide, respectively. The angled brackets in $\Iabs$ denote an averaging in the transverse directions ($xy$ plane), which translates the vector field $I_\mathrm{abs}(x,y)$ to an effective scalar quantity.

Assuming that the power launched into the waveguide mode is $P$ (in Watt), then the effective intensity from the mode-graphene overlap is $\Ieff=P/\AGeff$ (in W/m$^2$), where $\AGeff$ is an effective area defined, here, as
\begin{equation}
    \AGeff = 2 Z_0 \mathcal{P} \frac{\int_G|\fv{e}_\parallel(\omega;x,y)|^2d\ell}{\int_G|\fv{e}_\parallel(\omega;x,y)|^4d\ell}.
    \label{eq:AGeff}
\end{equation}
In this expression, and with reference to Fig.~\ref{fig:1:Schems}(b) axes, the complex vector $\fv{e}_\parallel(\omega;x,y)$ is the E-field component of the eigenmode that is parallel to graphene at frequency $\omega$, $\int_G d\ell$ denotes integration along the trace of the graphene sheet in the waveguide cross-section ($xy$ plane), $Z_0\approx377~\Omega$ is the free-space impedance, and $\mathcal{P}=0.5\Real{\int\fv{e}\times\fv{h}^*dxdy}$ (in Watt) is the eigenmode's power-normalization constant. The expression of Eq.~\eqref{eq:AGeff} is derived from the observation that graphene nonlinearities are proportional to the local E-field intensity tangential to graphene, $|\mathbf{E}_\parallel(x,y)|^2$, and power-weighted in the cross-section. Note that this $\AGeff$ is not equivalent to the effective area used in waveguides with bulk nonlinear media, such as silica-core optical fibers \cite{agrawal2012NLFO} or silicon (semiconductor) integrated structures \cite{Lin2007,Afshar2009,Daniel2010,Pitilakis2013}.

The effective area of Eq.~\eqref{eq:AGeff} is a characteristic of the waveguide mode in the linear regime and can be assumed unchanged in the nonlinear NIR regime too, as graphene's refractive effect (i.e., on the waveguide-mode concentration and profile) is perturbative across a large bandwidth around $\omega$. Then, using Eq.~\eqref{eq:AGeff}, \eqref{eq:Iabslocal}, and the $I=|E|^2/2Z_0$ assumption, we can finally define the spatiotemporal evolution of the effective absorbed power as
\begin{equation}
    \Iabs_{i/e}(z,t) = \Real{\sig{1}_{i/e}(z,t)} Z_0 \frac{P(z,t)}{\AGeff}.
    \label{eq:IabseffWG}
\end{equation}
This expression couples the pulse-envelope power $P(z,t)$ to graphene transient conductivity $\sig{1}_{i/e}(z,t)$, via the GHEM [e.g. Eqs.~\eqref{eq:PDE_SAM_Energy}, \eqref{eq:PDE_SAM_nPG}, \eqref{eq:mus_Electroneutrality} and the underlying Kubo formulas in the Appendix~\ref{app:sigmaCalc}] eventually distorting a high-power pulse as it propagates along the nonlinear waveguide. We reiterate that in the free-space case studied in \cite{Mikhailov2019HEM}, which is essentially a zero-dimensional problem, the absorbed intensity is given as $\Iabs=A I_\mathrm{eff,inc}$, where $I_\mathrm{eff,inc}$ is the effective incident optical intensity and $A$ is a power-absorption coefficient by Eq.~\eqref{eq:TinkhamAbsorption}; note that this is a spectral domain formula, so a temporal convolution is implied when studying pulsed excitation. Finally, we stress that diffraction-limited highly-confining nanophotonic waveguides can have exceedingly small $\AGeff\ll1~\mu$m$^2$ which corresponds to effective incident intensities $I_\mathrm{eff,inc}\gg1$~MW/cm$^2$, as inputs to the GHEM. For instance, the TE-polarized mode in the silicon-slot waveguide of \cite{Pitilakis2021} has $\AGeff<0.1~\mu$m$^2$, meaning that $\Ieff=1$~GW/cm$^2$ corresponds to a modest on-chip $P_\mathrm{peak}<1$~W.

\subsubsection{Intraband Rate Equation}

In the intraband rate equation, we note that the carrier temperature ($T$, common for hot electrons and holes) can be used instead of the total plasma energy. This is more efficient for transient numerical computations, as it does not require a $\{\mu_e,\mu_h,\Epd_T\}\rightarrow T$ mapping [i.e., an inversion of Eq.~\eqref{eq:Edeh}] to acquire the input parameters $\{\mu_e,\mu_h,T\}$ required by the Kubo formulas, Eqs.~(\ref{eq:s1i_full}-\ref{eq:s1e_full}), to extract the nonlinear $\sig{1}_{i,e}$ response. The rate equation for the carrier temperature can be easily deduced with the chain rule,
\begin{equation}
    \frac{\partial T}{\partial t} = \frac{\partial T}{\partial \Epd_T} \frac{\partial\Epd_T}{\partial t},
    \label{eq:ChainRule}
\end{equation}
and properties of the Fermi-Dirac integral derivatives, Eq.~\eqref{eq:FDIderiv}. The resulting rate equation for the carrier temperate, which replaces Eq.~\eqref{eq:PDE_SAM_Energy} in the GHEM equation systems, reads 
\begin{equation}
    \frac{\partial T}{\partial t} = \frac{T}{3\Epd_T-\mu_e n_e +\mu_h n_h}
       \bigg( \Iabs_i - \frac{\Epd_T-\Epd_\mathrm{QE}}{\tau_\Ec} \bigg),
    \label{eq:PDE_AP_Temp}
\end{equation}
where the derivation of the first factor in the RHS can be found in the Appendix~\ref{sec:Proof_EnergyTemp} and term in parentheses is actually the RHS term of the plasma energy density rate equation, Eq.~\eqref{eq:PDE_SAM_Energy}. We have verified the equivalence of the two equation systems.

\subsubsection{Interband Rate Equation}

Recent experiments \cite{Demongodin2019,Vermeulen2018} suggest that the density of the photogenerated carriers in graphene has an upper limit. This is in contrast to bulk semiconductors where the carrier density is much higher, even at thermal equilibrium; moreover, under high-intensity excitation, the recombination rate increases proportionally to the square of the carrier density, thus making a saturation density practically unreachable in bulk semiconductors. Now, in graphene, it is theorized that such a saturation density is indeed reachable and can be the source of non-saturable losses observed in optical and THz experiments \cite{Bao2009,Bao2010,Zhang2010}. The saturation density is expected to depend on the quality of the graphene sample (carrier mobility) and its equilibrium carrier density. Now, the rate Eq.~\eqref{eq:PDE_SAM_nPG} does not force a hard limit on $n_\mathrm{PG}$, except for any decrease of $\Real{\sig{1}_e}$ predicted by the Kubo formula, Eq.~\eqref{eq:s1e_integrable}, as the carrier temperature increases. However, this decrease is asymptotic which means that the GHEM can indeed produce unrealistic values of $n_\mathrm{PG}$ as the illuminating intensity increases. In order to avoid such a scenario, we heuristically introduce a saturation coefficient in the generation term of the rate equation, $G_\mathrm{sat}$, as follows:
\begin{equation}
    \frac{\partial n_\mathrm{PG}}{\partial t} = G_\mathrm{sat} \frac{\Iabs_e}{\hbar\omega} - \frac{n_\mathrm{PG}}{\tau_\mathrm{rec}} \bigg(1+\frac{n_\mathrm{PG}}{n_{T_0}}\bigg).
    \label{eq:PDE_AP_nPG_Gsat}
\end{equation}
The phenomenological coefficient $G_\mathrm{sat}$ is a monotonically decreasing function of $n_\mathrm{PG}$ and can be used to qualitatively explain experimental observations such as non-saturable conductivity or fit the GHEM to such results. More discussion in Section~\ref{sec:4:GHEMresults} (e.g., Fig.~\ref{fig:GHEM_CW_varNsat}), but unless explicitly stated assume $G_\mathrm{sat} \rightarrow 1$.

On a less critical note, the quadratic factor in the carrier recombination rate [the second term in the parentheses in the RHS of Eq.~\eqref{eq:PDE_SAM_nPG} or ~\eqref{eq:PDE_AP_nPG_Gsat}], which scales with the minority carrier density, can be dropped for $|\mu_c|>0.1$~eV, without deviations in the results. Nevertheless, the numerical burden is negligible, so we retain the term.

\section{Nonperturbative Regime: Results} \label{sec:4:GHEMresults}
In this section, we will present the results obtained using the GHEM developed in Section~\ref{sec:3:GHEMtheory} at photon energy $\hbar\omega=0.8$~eV ($\lambda_0=1550$~nm). This model consists of six unknown graphene variables, $\{T,n_\mathrm{PG},\mu_e,\mu_h,\mu_{e0},\mu_{h0}\}$, and six equations: \eqref{eq:PDE_AP_Temp}, \eqref{eq:PDE_AP_nPG_Gsat} and the set of four electroneutrality Eqs.~\eqref{eq:mus_Electroneutrality}. Moreover, as we are presently studying only graphene's nonlinear photoconductivity, we assume that the instantaneous effective absorbed intensity is related to the incident E-field amplitude with
\begin{equation}
    \Iabs_{i,e}=\frac{1}{2}\Real{\sig{1}_{i,e}}|E|_\mathrm{eff}^2=\Real{\sig{1}_{i,e}} Z_0 I_\mathrm{eff},
\end{equation}
i.e., as if we were in a short (unitary length) waveguide where the modal electric field that is tangential to graphene has an effective intensity of $I_\mathrm{eff}=|E|_\mathrm{eff}^2/(2 Z_0)$, in W/m$^2$. This is not equivalent to the case studied in \cite{Mikhailov2019HEM}, where there is additionally the full conductivity $\sig{1}$ in the denominator of Eq.\eqref{eq:Iabs_SAM_Tinkham}, but nevertheless the same trends will qualitatively hold as the generation rates are primarily proportional to $\Real{\sig{1}_{i,e}}$, in both cases. According to the analysis in Section~\ref{sec:Iabseff}, given the instantaneous power $P$ (in Watt) launched in the waveguide, we can calculate the effective E-field intensity that feeds the GHEM as $I_\mathrm{eff}=P/\AGeff$ (in W/m$^2$) using Eq.~\eqref{eq:AGeff} for $\AGeff$. The term \textit{photoconductivity}, as used here, refers to the optically induced complex-valued change in graphene's surface conductivity:
\begin{equation}
    \Delta\sig{1}(t) = \sig{1}_\mathrm{GHEM}(t) - \sig{1}_\mathrm{lin},
    \label{eq:photoncond}
\end{equation}
where $\sig{1}_\mathrm{lin}$ is the surface conductivity at the absence of high-power excitation (linear regime) and $\sig{1}_\mathrm{GHEM}(t)$ is the nonlinear conductivity predicted by the GHEM under high-power excitation, which can moreover be time-dependent. Equation~\eqref{eq:photoncond} can also be normalized by the universal conductivity $\sigma_0$. Finally, unless otherwise stated, we use a momentum relaxation rate with energy dependence for the intraband $\Gamma_i$ (refer to Appendix~\ref{sec:s1intra} for details) and a constant $\Gamma_e=0.5$~meV ($\tau_e=1.3$~ps) for the interband mechanism.

\subsection{Quasi-Perturbative Regime}

In this regime, the illuminating intensity produces a nonlinear response through the GHEM, i.e., a photoconductivity $\Delta\sig{1}$, but this can be considered as a small perturbation compared to the linear (low-intensity) conductivity $\sig{1}_\mathrm{lin}$. In this case, we can use a Taylor approximation around the effective illuminating (CW) intensity $|E|_\mathrm{eff}^2=2 Z_0 \Ieff$ to extract a quasi-perturbative $\sig{3}_{xxxx}(-\omega_0,+\omega_0,+\omega_0)$ for the self-acting Kerr-type nonlinearity \cite{Alexander2018},
\begin{equation}
    \sig{3}_\mathrm{QP} = \frac{2}{3} \frac{\partial\sig{1}_\mathrm{GHEM}}{\partial|E|_\mathrm{eff}^2}.
    \label{eq:s3_QuasiPertu}
\end{equation}
Note that this approximation is applicable up to a threshold value of $\Ieff$, for which the condition $|\Delta\sig{1}|\ll|\sig{1}_\mathrm{lin}|$ holds. For the $|\mu_c|<\hbar\omega/2$ cases, the threshold intensity for this regime was found to be in the order of 1~MW/cm$^2$ (few tens of mW on-chip power), whereas it was much higher for $|\mu_c|>\hbar\omega/2$ where nonlinear response practically vanishes. Note that $\sig{3}_\mathrm{QP}$ is reduced, in both real and imaginary part, when that intensity threshold is surpassed.

\begin{figure}[h]
    \centering
    \includegraphics[width=85mm]{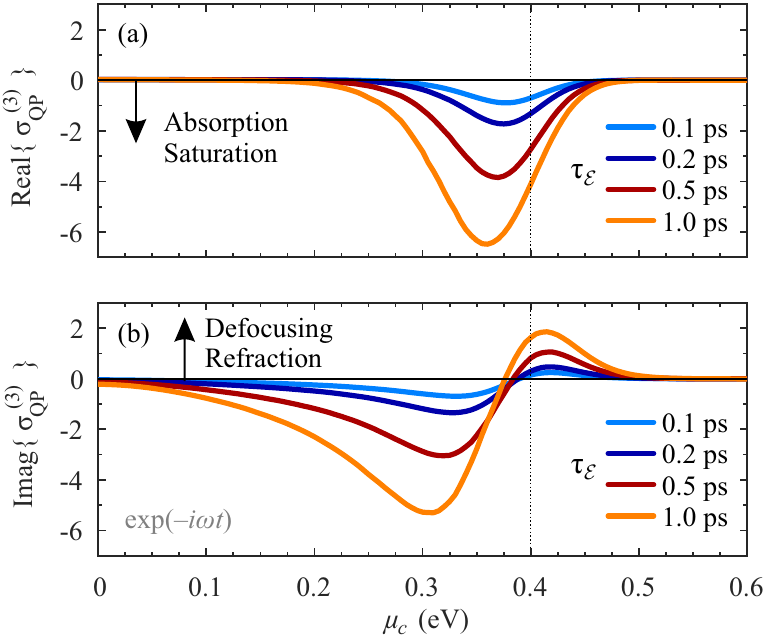}
    \caption{Effect of phenomenological energy relaxation lifetime $\tau_\varepsilon$ on the quasi-perturbatively calculated $\sig{3}_\mathrm{QP}$ at $\lambda_0=1550$~nm. (a) Real and (b) imaginary parts of the $\mu_c$ spectra in units of $10^{-19}$~S(m/V)$^2$.}
    \label{fig:s3vsmu_QuasiPert_varTauEps}    
\end{figure}

In Fig.~\ref{fig:s3vsmu_QuasiPert_varTauEps} we present the quasi-perturbative $\mu_c$ spectra for $\sig{3}_\mathrm{QP}$, extracted from Eq.~\eqref{eq:s3_QuasiPertu}, for a few phenomenological energy relaxation lifetimes $\tau_\Ec$. The rest of the GHEM parameters take their default values, i.e., $T=300$~K, $\lambda_0=1550$~nm (0.8~eV), $\Gamma_i=\Gamma_i(\Ec;\zeta,\Eimp)$ with $\{\zeta,\Eimp\}=\{4,30~\mathrm{meV}\}$, $\Gamma_e = 0.5$~meV, and $\tau_\mathrm{rec}=10\tau_\Ec$. Evidently, when $\tau_\Ec$ increases, the nonlinearity is in overall larger, owing to the prolonged thermalized state of the plasma. Moreover, we observe a qualitative agreement with the perturbative regime, Fig.~\ref{fig:s3vsmu_varGi}, i.e., a peaking of defocusing nonlinearity when the chemical potential is tuned near the half-photon energy and a saturable absorption regime near and below that energy. However, there are also some important differences in these spectra: First and foremost we observe much higher peak values of $\sig{3}_\mathrm{QP}$, in the order to $10^{-19}$~S(m/V)$^2$ for both the real and imaginary parts, even for low $\tau_\Ec$ values. Secondly, $\Real{\sig{3}_\mathrm{QP}}$ vanishes except near and below half-photon energy, whereas in Fig.~\ref{fig:s3vsmu_varGi} it had a flat value from $\mu_c=0$ up until the half-photon energy. Both these features are in accordance with the experimental observations in \cite{Alexander2018} [and Fig.~S6(a-b) in its supporting information], where a simpler GHEM was used, neglecting the interband mechanism and the photogenerated carrier density. Thirdly, we observe a peaking of self-focusing refractive nonlinearity ($\Imag{\sig{3}_\mathrm{QP}}<0$) below half-photon energy, which is moreover three times higher than the subsequent defocusing peak on half-photon energy. This interesting new feature emerges independently of other GHEM parameters, i.e., momentum relaxation rates ($\Gamma_{i,e}$) or operating wavelength, and appeared neither in Fig.~\ref{fig:s3vsmu_varGi} nor in \cite{Alexander2018}; we attribute it to the interband absorption mechanism and to the non-negligible $n_\mathrm{PG}>0.15n_{T_0}$ it generates. In any case, this feature indicates a high sensitivity (sign flip) of the refractive nonlinearity on $\mu_c$, i.e., on electrical tuning. 

\subsection{Nonperturbative Static Response}

When the illuminating power increases beyond the threshold discussed in the previous section, graphene photoconductivity transcends the perturbative regime. In terms of the GHEM parameters, this is reflected in a non-negligible carrier temperature ($\Delta T/T_0>1$\%) and density ($n_\mathrm{PG}/n_{T_0}>1$\%) increase, as well as a shift in the `dominant' quasi-Fermi level (e.g. in $\mu_h$ if $\mu_c<0$ ). In this non-perturbative regime, we cannot use the third-order effect formalism, and directly study the photoconductivity $\Delta\sig{1}$ as a function of graphene parameters and the effective incident CW intensity. 

We focus on the NIR spectral region, and specifically to the telecom C-band wavelength $\lambda_0=1550$~nm ($\hbar\omega\approx0.8$~eV). In this region, graphene's response is mostly absorptive and does not contribute to waveguiding, i.e., the contribution of $\Imag{\sig{1}}$ is negligible compared to index guiding. Moreover, for chemical potential well above $\hbar\omega/2$ graphene is practically transparent and a mild induced absorption is expected when the illumination intensity increases \cite{Mikhailov2019HEM}, so we restrict our analysis to $\mu_c<0.5$~eV, i.e., just above the half-photon energy. 

An important conclusion drawn directly from \cite{Mikhailov2019HEM}, Eq.~(47) therein, is that the nonlinear threshold (NLT) in the intensity is inversely proportional to the phenomenological lifetimes $\tau_\Ec$ and $\tau_\mathrm{rec}$, governing the intraband energy relaxation and interband recombination, respectively. Secondarily, in the absence of a carrier-saturation density [$G_\mathrm{sat}=1$ in Eq.~\eqref{eq:PDE_AP_nPG_Gsat}], the nonlinear response depends only on the ratio $\tau_\Ec/\tau_\mathrm{rec}$ which should moreover be smaller than unity to validate the assumptions made. In this work, we assume values $\tau_\Ec=1$~ps and $\tau_\mathrm{rec}=10$~ps. 

\begin{figure}[h]
    \centering
    \includegraphics[width=160mm]{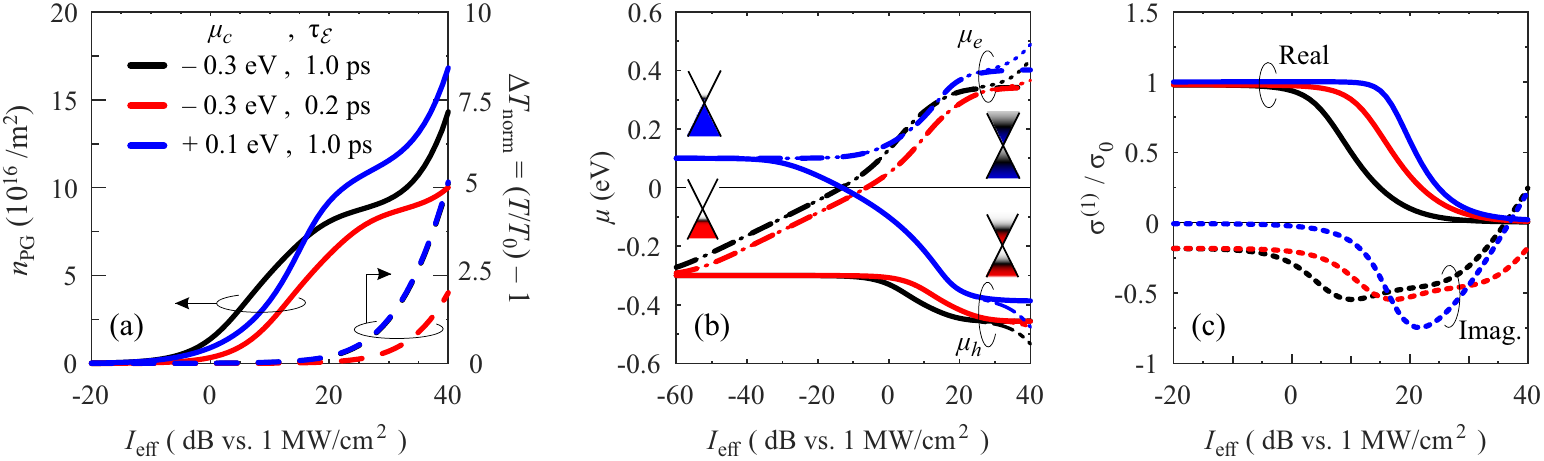}
    \caption{Saturation curves for (a) carrier temperature and $n_\mathrm{PG}$, (b) out-of-equilibrium and quasi-equilibrium chemical potentials, and (c) surface conductivity. The three colors denote three different combinations of $\{\mu_c,\tau_\Ec\}$. The insets in panel (b) depict band-filling as predicted by the GHEM at very low and very high intensities, for two of the combinations. In all cases: $T_0=300$~K, $\lambda_0=1550$~nm and $\tau_\mathrm{rec}=10\tau_\Ec$.}
    \label{fig:GHEM_CW_6par_s1}    
\end{figure}

In Fig.~\ref{fig:GHEM_CW_6par_s1}, we extract the \textit{saturation curves} for the six graphene variables, $\{T,n_\mathrm{PG},\mu_e,\mu_h,\mu_{e0},\mu_{h0}\}$, and the induced conductivity as a function of effective incident intensity, $\Ieff$. We consider three combinations of $\{\mu_c,\tau_\Ec\}$, i.e., equilibrium chemical potential and energy relaxation lifetime, respectively. In panel (a), we observe the exponential increase in carrier density and temperature with the logarithm of intensity. Also, the NLT for carrier photogeneration onset is almost a hundred times lower than the NLT for temperature increase; this is anticipated as for low $|\mu_c|$ the interband mechanism dominates the total conductivity. Moreover, we observe that decreasing $\tau_\Ec$ (energy relaxation lifetime) increases the NLT and that the temperature increase is indistinguishable for $|\mu_c|=0.1$ and 0.3~eV (black and blue dashed curves overlap). In panel (b), we see that the minority quasi-Fermi level (e.g., $\mu_e$ if $\mu_c<0$) has a very low NLT and experiences a sign-flip as $n_\mathrm{PG}$ increases. Adversely, the dominant quasi-Fermi level magnitude increases thus leading to a large $\Delta\mu_{e-h}$ that approaches $\hbar\omega$. Finally, the quasi-equilibrium (QE) chemical potentials are plotted with thin dot/dash-dot curves, which overlap with the corresponding `hot' chemical potentials and only deviate (to higher magnitudes) at very high intensities. Finally, in panel (c), we present the conductivity clearly showing deep SA beyond the NLT of $10$~MW/cm$^2$; the NLT in this case is what is phenomenologically referred to as \textit{saturation intensity}, $I_\mathrm{sat}$, i.e., the intensity where absorption drops to half of its low-power (linear-regime) value. 

It is worth pointing out that both real and imaginary parts of $\sig{1}$ in Fig.~\ref{fig:GHEM_CW_6par_s1}(c) bear a qualitative resemblance to the corresponding $\mu_c$-dependence, Fig.~\ref{fig:s1Linear}: the negative peaking of the imaginary part coincides with the halving of the real part. This resemblance unveils an association between the logarithm of intensity in the nonlinear regime and the chemical potential (or square-root of carrier density) in the linear regime. 

\begin{figure}[h]
    \centering
    \includegraphics[width=160mm]{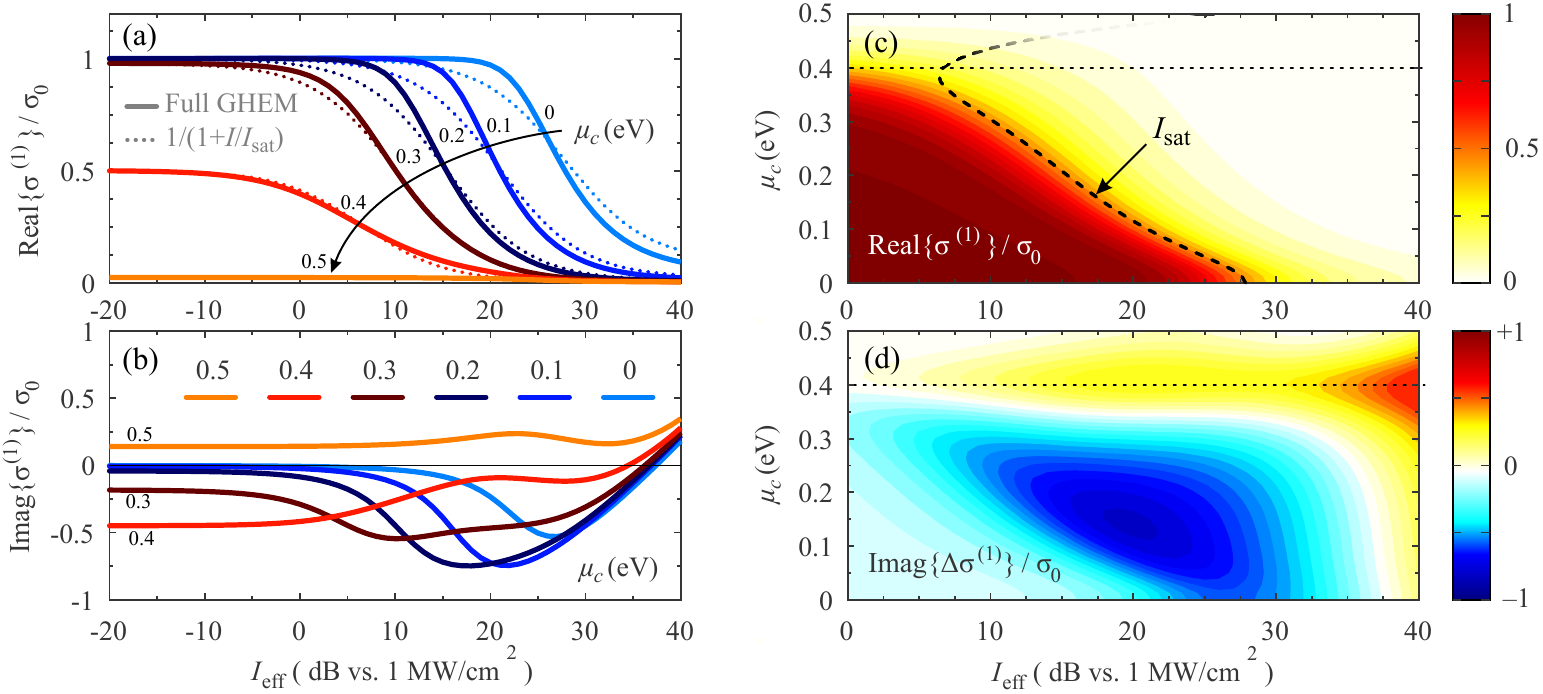}
    \caption{(a) Real and (b) imaginary surface conductivity vs. effective incident intensity, $\Ieff$, for a few $\mu_c$ values. (c) Real conductivity and (d) imaginary photoconductivity, ${\Delta\sig{1}}$, vs. $\{\Ieff,\mu_c\}$. In all cases: $T_0=300$~K, $\lambda_0=1550$~nm, $\tau_\Ec=1$~ps, and $\tau_\mathrm{rec}=10$~ps.}
    \label{fig:GHEM_CW_varMu}    
\end{figure}

Figure~\ref{fig:GHEM_CW_varMu} is devoted to the effect of the voltage-tunable equilibrium chemical potential and illuminating intensity on GHEM-predicted nonlinear conductivity. Panels (a)-(b) present saturation curves for a few values of $\mu_c$ whereas panels (c)-(d) contain heatmaps with finer resolution. The conclusions drawn from these results are: (i) The empirical fit for SA, $\Real{\sig{1}(I)} = \sig{1}_{ns} + \Delta\sig{1}_\mathrm{sat}/(1+I/I_\mathrm{sat})$, where $\sig{1}_{ns}$ is the non-saturable conductivity and $\Delta\sig{1}_\mathrm{sat}=\Real{\sig{1}(0)}-\sig{1}_{ns}$, agrees with the full-GHEM curves; (ii) The saturation intensity, black curve labeled $I_\mathrm{sat}$ in panel (c), decreases over a hundredfold as $\mu_c=0\rightarrow\hbar\omega/2$ and then increases above that; (iii) The refractive part of the photoconductivity, $\Imag{\Delta\sig{1}}$, is negative except near half-photon energy, where is goes positive.

\begin{figure}[h]
    \centering
    \includegraphics[width=150mm]{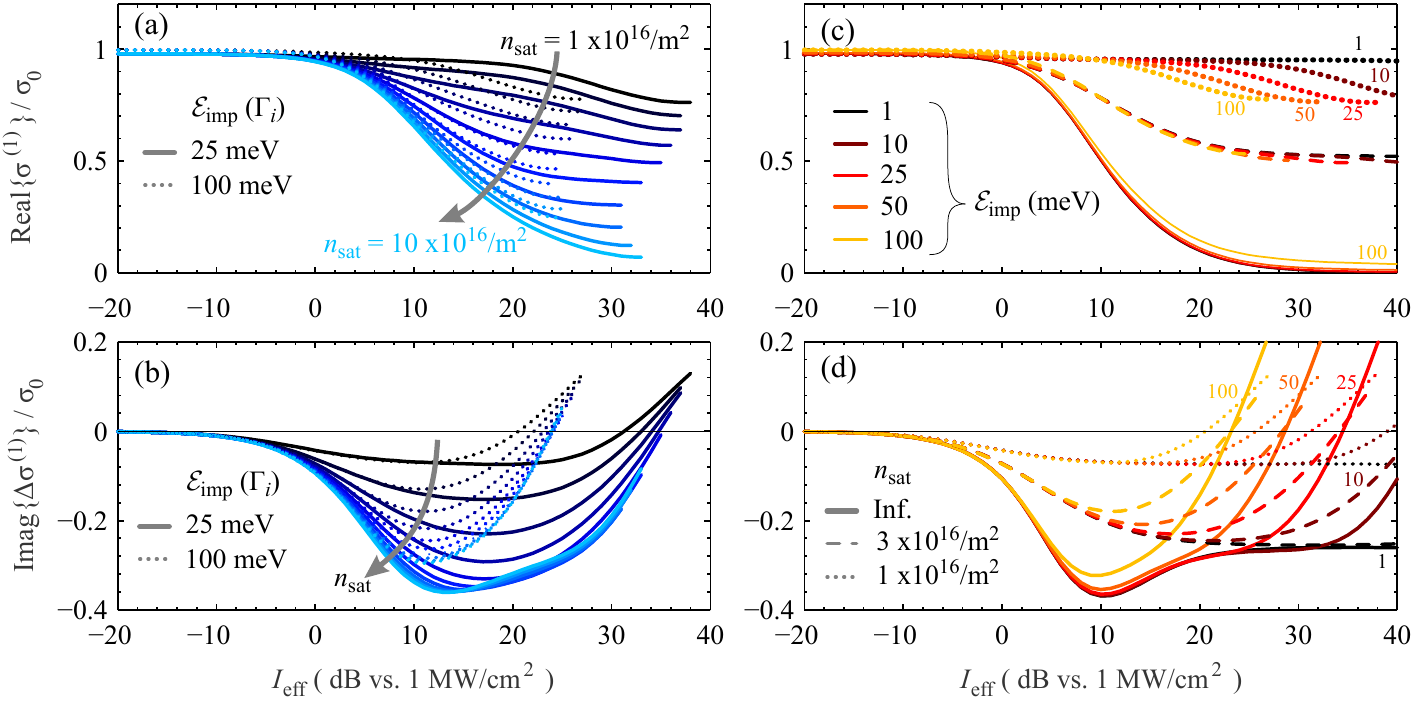}
    \caption{Effect of the saturation carrier density, $n_\mathrm{sat}$, and the energy of Coulomb impurities, $\Eimp$, on the (a,c) real and (b,d) imaginary parts of graphene surface conductivity. In panels (a,b) curve color and line-style denote different $n_\mathrm{sat}$ and $\Eimp$, respectively, while the convention is opposite in panels (c,d). In all cases: $|\mu_c|=0.3$~eV, $T_0=300$~K, $\lambda_0=1550$~nm, $\tau_\Ec=1$~ps, and $\tau_\mathrm{rec}=10$~ps.}
    \label{fig:GHEM_CW_varNsat}    
\end{figure}

Closing the study of the steady-state solutions of the GHEM, we present the saturation-carrier density effect on the photoconductivity. We employ the simple empirical form $G_\mathrm{sat}(n_\mathrm{PG})=(1-n_\mathrm{PG}/n_\mathrm{sat})$ for the saturation factor in the generation term of Eq.~\eqref{eq:PDE_AP_nPG_Gsat}, as proposed in \cite{Vermeulen2018}. $n_\mathrm{sat}$ is the saturation carrier density that $n_\mathrm{PG}$ cannot surpass, which would be proportional to the equilibrium carrier density, i.e., in the order of few $10^{16}$/m$^2$ for the $\mu_c<\hbar\omega/2$ studied in this work. In parallel to $n_\mathrm{sat}$, we also study the effect of the monolayer quality through the equivalent Coulomb-impurity energy, $\Eimp$, which governs the energy-dependence of the intraband momentum relaxation lifetime, $\Gamma_i$; for more details refer to Appendix~\ref{sec:s1intra} and Eq.~\eqref{eq:Gintra}.

From Fig.~\ref{fig:GHEM_CW_varNsat}(a,b) we see that $n_\mathrm{sat}>10^{16}$/m$^2$ results in an overall quenching of the photoconductivity. The residual $\Real{\sig{1}}$ at high intensities can be fit to non-saturable losses observed in recent experiments \cite{Demongodin2019} and thus estimate $n_\mathrm{sat}$. Concerning $\Imag{\sig{1}}$, the carrier-density saturation seems to quench only the negative feature that peaks near $I_\mathrm{sat}$, leaving the positive up-shooting mostly unimpeded. This nonlinear refraction trend is also in line with recent experiments \cite{Vermeulen2018,CastellLurbe2020}, where spectral broadening of negatively chirped pulses was associated with $\Imag{\sig{1}}>0$ (under the $e^{-i\omega t}$ phase-convention) and moreover scaling with the square-root of the carrier density. Finally, note that some of the curves in Fig.~\ref{fig:GHEM_CW_varNsat} do not extend all the way to $+40$~dB. We found an unphysical re-rising of $\Real{\sig{1}}$ after a threshold intensity, coinciding with a significant deviation of the QE chemical potentials from their corresponding `hot' counterparts. This anomaly can be corrected by increasing the lattice temperature ($T_0$) after the real part ceases to decrease, which leads us to believe that photothermal effects must be accounted for in very high CW intensities, to further extend the validity range of the GHEM; refer to relevant discussion in Section~\ref{sec:Outlook}. Finally, we investigate the effect of monolayer sample quality through $\Eimp$, whose lower values correspond to higher carrier mobility, hence higher quality. From Fig.~\ref{fig:GHEM_CW_varNsat}(c), we observe that its effect is meaningful only for very low $n_\mathrm{sat}$, where lower quality corresponds to lower $I_\mathrm{sat}$; note that for $\Eimp\rightarrow0$, the SA vanishes. The effect of sample quality is particularly pronounced on the refractive part of the surface conductivity, Fig.~\ref{fig:GHEM_CW_varNsat}(b,d): An increase in $\Eimp$ (quality decrease) contributes to orders of magnitude decrease in the intensity threshold where $\Imag{\Delta\sig{1}}$ crosses from negative to positive. Evidently, this graphene parameter can have significant impact on applications relying on refractive nonlinearities, e.g., pulse shaping, spectral broadening, etc.

\subsection{Nonperturbative Transient Response}

\begin{figure}[h]
    \centering
    \includegraphics[width=160mm]{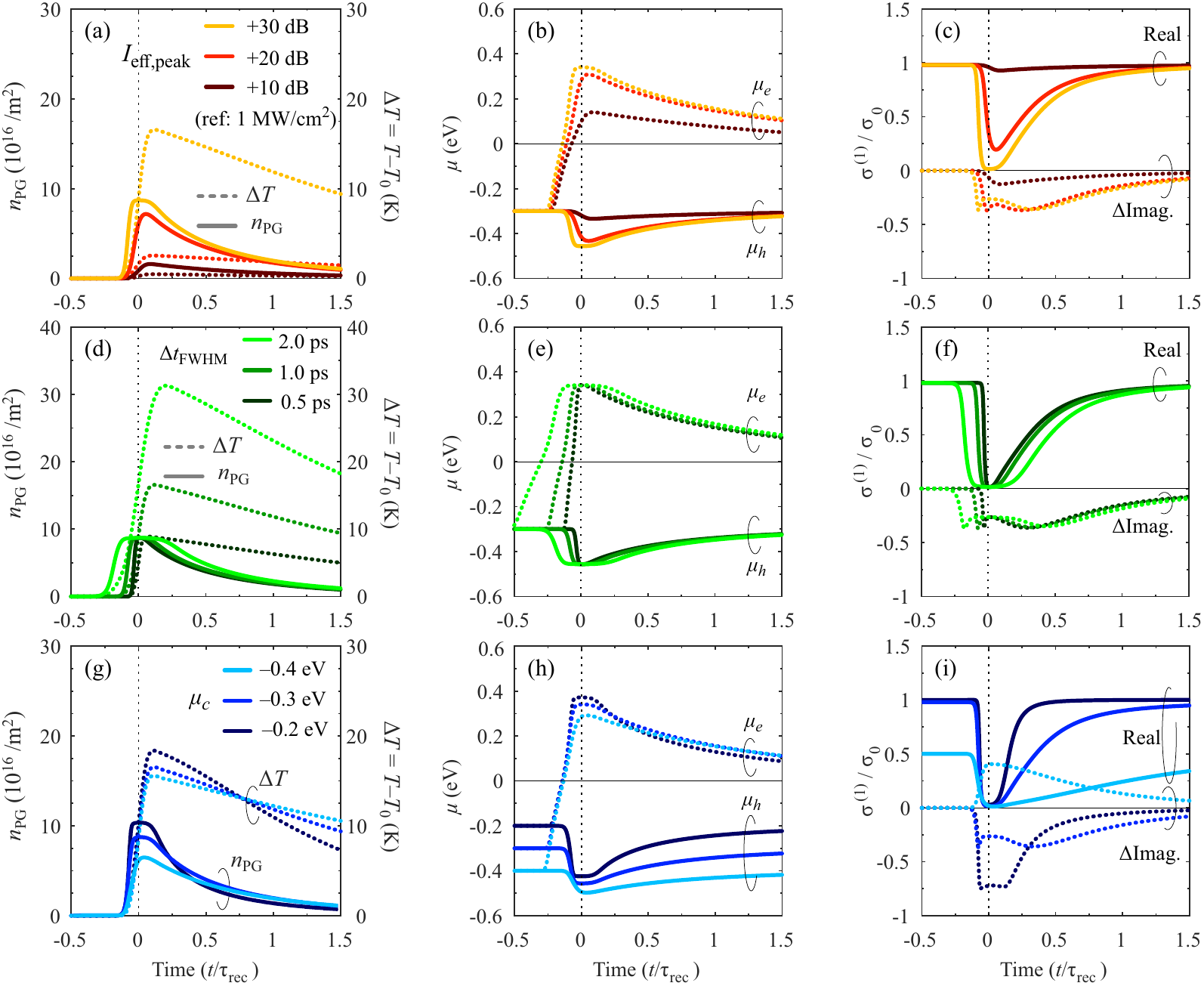}
    \caption{Transient GHEM response of (a,d,g) photogenerated carrier density and temperature shift, (b,e,h) quasi Fermi levels, (c,f,i) surface conductivity. We vary (a,b,c) the impinging intensity, (d,e,f) the pulse duration, and (g,h,i) the equilibrium chemical potential; default values are $I_\mathrm{eff,peak}=+30$~dB vs. $1$~MW/cm$^2$, $\Delta t_\mathrm{FWHM}=1$~ps, and $\mu_c=-0.3$~eV, respectively. In all cases $T_0=300$~K, $\lambda_0=1550$~nm, $\tau_\Ec=1$~ps, and $\tau_\mathrm{rec}=10$~ps, to which the time axis is normalized.}
    \label{fig:GHEM_Transient_VarAll}    
\end{figure}

We now proceed to the transient solution of the GHEM equations (refer to Appendix~\ref{app:Numerical_Tips} for implementation details) in the retarded time-frame of a Gaussian-enveloped ps-pulse, with effective intensity $\Ieff(t)=\exp[-2 c_{SG} (t/\Delta t_\mathrm{FHWM})^{2N_{SG}}]$, where $c_{SG}=2^{(2N_{SG}-1)}\ln{2}$ is a duration normalization constant (equal to 1.3863 for regular Gaussian pulse of order $N_{SG}=1$) and $\Delta t_\mathrm{FWHM}$ is the full-width at half-maximum of the pulse power. We hereby investigate the effect of the controllable GHEM parameters: In Fig.~\ref{fig:GHEM_Transient_VarAll} panels (a)-(c), the effect of pulse peak intensity $I_\mathrm{eff,peak}$, in panels (d)-(f), the effect of $\Delta t_\mathrm{FWHM}$, and, in panels (g)-(i), the effect of chemical potential at the absence of illumination, $\mu_c$. When not varied, default values are $I_\mathrm{eff,peak}=+30$~dB vs. $1$~MW/cm$^2$, $\Delta t_\mathrm{FWHM}=1$~ps, and $\mu_c=-0.3$~eV. The columns-of-panels of Fig.~\ref{fig:GHEM_Transient_VarAll} correspond to the ones in Fig.~\ref{fig:GHEM_CW_6par_s1}: panels (a,d,g) depict $\{n_{GP},\Delta T\}$ vs. time, panels (b,e,h) depict $\mu_{e,h}$ vs. time, and panels (c,f,i) depict $\Real{\sig{1}}$ and $\Imag{\Delta\sig{1}}$ vs. time. In all cases $T_0=300$~K, $\lambda_0=1550$~nm, $\tau_\Ec=1$~ps, and $\tau_\mathrm{rec}=10$~ps, to which the time axis is normalized. 

Analyzing the results in Fig.~\ref{fig:GHEM_Transient_VarAll}, some interesting conclusions can be drawn, valid for all cases, i.e., when varying either $I_\mathrm{eff,peak}$, $\Delta t_\mathrm{FWHM}$ or $\mu_c$: (i) The nonlinear photoconductivity correlates best with $n_\mathrm{PG}$, rather than $T$ or $\mu$, implying that this GHEM parameter is the most important to qualitatively capture the temporal dynamics. (ii) The $\Imag{\Delta\sig{1}}$ shows some rippling around the pulse peak, as expected by the non-monotonic CW curves as intensity increases, Fig.~\ref{fig:GHEM_CW_6par_s1}(c); this rippling would impart a chirp on a pulse that propagates along the monolayer. (iii) The carrier temperature decays much slower than $n_\mathrm{PG}$, despite $\tau_\Ec=0.1\tau_\mathrm{rec}$; this apparent contradiction is due to the factor $\partial T/\partial \Epd_T$ in the RHS of Eq.~ \eqref{eq:PDE_AP_Temp} [and Eq.~\eqref{eq:dEnergy_dTemp}] that increases the lifetime. We performed a consistency check on the quantity $\Epd_T(t)-\Epd_\mathrm{QE}$, as in Eq.~\eqref{eq:PDE_SAM_Energy} to which the faster rate $\tau_\Ec^{-1}$ is applied, which indeed decays faster than $n_\mathrm{PG}$. 

Some more interesting observations can be extracted from Fig.~\ref{fig:GHEM_Transient_VarAll}: In panel (d), we observe that peak $\Delta T$ is proportional to pulse duration, but peak $n_\mathrm{PG}$ is not affected in this range of pulse duration, $\Delta t_\mathrm{FWHM}=0.5$-$2$~ps. From panel (g), we observe that peak absolute $n_\mathrm{PG}$ increases with decreasing $|\mu_c|$, which consequently means that the difference in the normalized $n_\mathrm{PG}/n_{T_0}$ would be even higher; this apparent contradiction can be explained from Fig.~\ref{fig:GHEM_CW_6par_s1}(a), comparing the $n_\mathrm{PG}$ curves for $\mu_c=-0.3$ and $+0.1$~eV at the $+30$~dB power level considered here. Finally, in panel (i), we note the only case of $\Imag{\Delta\sig{1}}>0$, which happens for $|\mu_c|=\hbar\omega/2=0.4$~eV as expected from Fig.~\ref{fig:GHEM_CW_varMu}(d).

\begin{figure}[h]
    \centering
    \includegraphics[width=160mm]{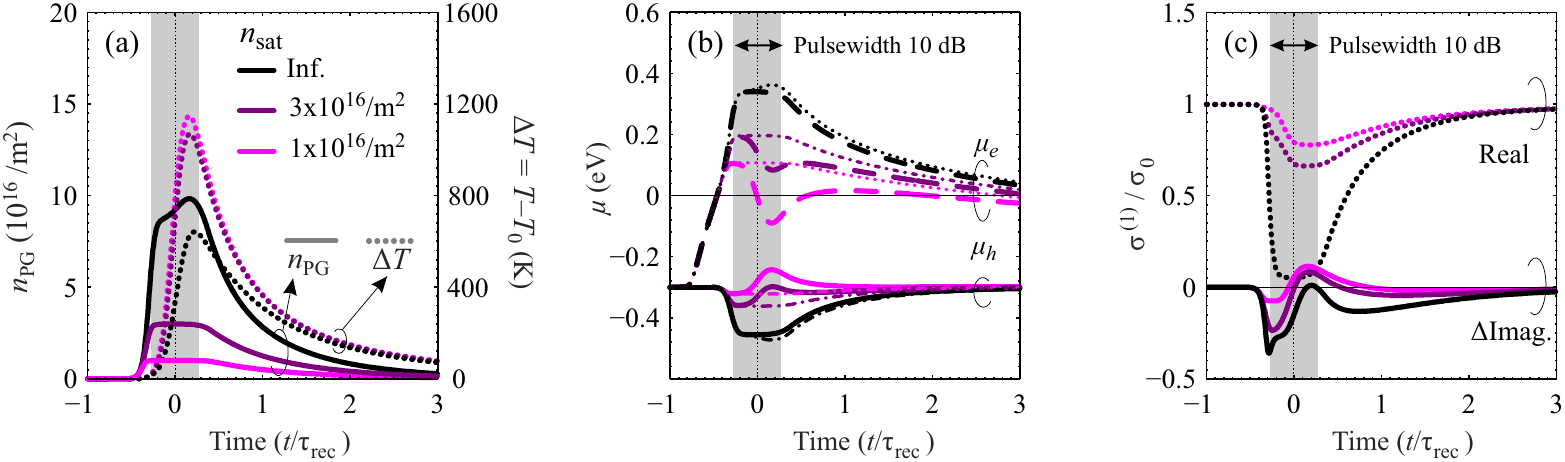}
    \caption{Effect of saturation carrier density $n_\mathrm{sat}$ on (a) $n_\mathrm{PG}$ and $\Delta T$, (b) $\mu_{e,h}$ and $\mu_{e_0,h_0}$, with thick and thin curves, respectively, and (c) $\sig{1}/\sigma_0$. Default GHEM parameters, i.e., as in Fig.~\ref{fig:GHEM_Transient_VarAll}, except for $\Delta t_\mathrm{FWHM}=3$~ps and $\Eimp=100$~meV, which corresponds to lower quality monolayer.}
    \label{fig:GHEM_Transient_VarNsat}     
\end{figure}

We now proceed to the study of the effect of the saturation carrier density, $n_\mathrm{sat}$, on the transient GHEM variables. We now consider a lower quality monolayer, $\Eimp=100$~meV corresponding to carrier mobility of about 630~cm$^2$/(Vs) on an air/oxide interface, and a longer Gaussian pulse duration of $\Delta t_\mathrm{FWHM}=3$~ps. Figure~\ref{fig:GHEM_Transient_VarNsat}(a) shows that $n_\mathrm{PG}\rightarrow n_\mathrm{sat}$ in all cases on pulse-peak and that the intraband absorption caused by the high impurity density leads to a large increase in carrier temperature, over 1000~K.  In panel (b), we see that the rippling of the quasi Fermi levels on pulse-peak translates to a similar rippling in the refractive (imaginary) part of the surface conductivity, panel (c), where we also observe that low $n_\mathrm{sat}$ values dramatically quench the attainable SA. In panel (b) we also plot the quasi Fermi levels at the QE (quasi-equilibrium) state, $\mu_{e_0,h_0}$, with thin dot/dash-dot lines; these differ visibly from the corresponding hot potentials and have a slower response.

\begin{figure}[h]
    \centering
    \includegraphics[width=85mm]{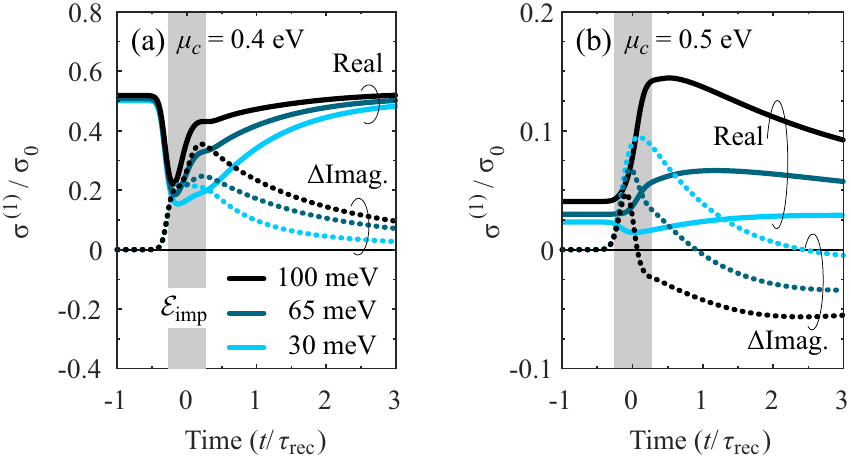}
    \caption{Effect of impurity density $\Eimp$ on surface photoconductivity, for monolayers biased at (a) $\mu_c=\hbar\omega/2$ and (b) $\mu_c=\hbar\omega/1.6$. Default GHEM parameters, i.e., as in Fig.~\ref{fig:GHEM_Transient_VarNsat}, and $n_\mathrm{sat}=3\times10^{16}$/m$^2$. The shaded area denotes the full width at 10~dB ($\Ieff(t)>100$~MW/cm$^2$).}
    \label{fig:GHEM_Transient_VarEimp}    
\end{figure}

Finally, in Fig.~\ref{fig:GHEM_Transient_VarEimp}, we investigate the effect of the monolayer impurity energy ($\Eimp\propto \sqrt{N_\mathrm{imp}}$, where $N_\mathrm{imp}$ is the effective impurity density) on the photoconductivity, when the equilibrium chemical potential is biased, panel (a), exactly on half-photon energy or, panel (b), just above it; $n_\mathrm{sat}=3\times10^{16}$/m$^2$ and the rest of the GHEM parameters are the same as in Fig.~\ref{fig:GHEM_Transient_VarNsat}. As expected from the static response curves, Fig.~\ref{fig:GHEM_CW_varNsat}, for $\mu_c=0.4$~eV we get $\Imag{\Delta\sig{1}}>0$, whose magnitude moreover increases with $\Eimp$, i.e., as the quality of the monolayer decreases; this feature is predominantly due to the intraband mechanism and the high thermalization that the carrier-density saturation brings. For the $\mu_c>\hbar\omega/2$ case, the overall photoconductivity (nonlinearity) is lower: the refractive part exhibits a positive peak followed by a sign-change with a long trailing negative $\Imag{\Delta\sig{1}}$, whereas the real part shows induced absorption for high impurity densities.

\section{Discussion} \label{sec:5:Discussion}
\subsection{Implications for Photonic Waveguides}

Having extensively analyzed the GHEM response per unit length of the waveguide one naturally wonders how the transient response evolves as the pulse travels down a waveguide. This effect can be modeled in terms of the nonlinear Schr\"{o}dinger equation (NLSE) framework, and numerically solved using the split-step Fourier method (SSFM); brief introduction for both can be found in Appendix~\ref{app:NLSE_SSFM}. 

With these remarks in mind, summarizing the GHEM photoconductivity results from Section~\ref{sec:4:GHEMresults}, i.e., for $\mu_c$ near and slightly below $\hbar\omega/2$, we conclude that graphene exhibits saturable absorption (SA), $\Real{\Delta\sig{1}}<0$, and positive self-focusing refraction, $\Imag{\Delta\sig{1}}<0$, for low and medium powers. Then, above an intensity threshold, the refraction switches to negative (defocusing), which moreover coincides with deep SA. The increasing quality (carrier mobility) of graphene was found to increase the SA depth, together with the intensity threshold for refraction sign-flipping. Forcing an upper boundary to photogenerated carrier density decreases the magnitude of nonlinearity (SA and refraction), without affecting the threshold intensities. Finally, the model predicts a strong coupling between the absorptive and refractive nonlinear regimes, with nontrivial boundaries between them.

Consequently, depending on the nonlinear application targeted (i.e., whether it relies on the absorptive or refractive part of $\sig{1}$ -- or to both), the graphene-related parameters must be judiciously chosen; these mainly include the graphene quality, its equilibrium chemical potential, and the pulse peak power. The dependence of graphene's nonlinear response on these parameters is non-trivial and often non-monotonic, meaning that optimal conditions might arise in banded windows within the parameter space. For instance: absorption modulation (e.g. for extinction-ratio improvement of an on/off-keying modulated bitstream) requires deep loss contrast, i.e., deep SA with a steep slope in the $\Real{\sig{1}}$-intensity curve; spectral broadening requires low losses (or deep SA) together with high $|\Imag{\sig{1}}|$; bright-soliton formation requires low losses (or deep SA) together with $\Real{n_2} \propto \Imag{\sig{1}}$ of opposite sign to the group velocity dispersion coefficient ($\beta_2$). In most of these applications, the optimal chemical potential would be close to half-photon energy, while the actual value (above or below it) would stem from specific compromises between its coupled refractive and absorptive parts.

\subsection{Qualitative Trends and Comparison to Experiments}

The dominant nonlinearity in the $\Real{\Delta\sig{1}}$ is SA, with a minimal $I_\mathrm{sat}<10$~MW/cm$^2$, found slightly below half-photon energy. This observation qualitatively agrees both with free-space experiments and with pulsed waveguide measurements \cite{Demongodin2019}. In the latter, a relatively shallow SA was measured, corresponding to high non-saturable (residual) conductivity, with negligible refractive nonlinearity. This behaviour can be explained by a combination of low carrier mobility and/or low carrier-saturation density in the framework of this GHEM. Moreover, fitting of the model parameters to experimentally characterized samples could also improve quantitative accuracy when optimizing such devices.

The refractive part of the nonlinear photoconductivity, $\Imag{\Delta\sig{1}}$, constitutes a rather interesting and relatively unexplored regime. Our GHEM qualitatively agrees with recent experiments in silicon and silicon nitride graphene-comprising waveguides \cite{Vermeulen2016,Vermeulen2018}, i.e., it predicts a positive $\Imag{\sig{1}}$ regime, which increases with carrier density. Concerning the quantitative comparison to the phenomenological model proposed in \cite{Vermeulen2018} and theoretically elaborated in \cite{CastellLurbe2020}, our model predicts a larger threshold stemming from a different dependence on the carrier density. Specifically, \cite{Vermeulen2018} proposes a direct proportionality $\Imag{\Delta\sig{1}}\propto\sqrt{n_T}$ in order to explain the superlinear (exponential) rising in spectral-broadening that was measured, as the waveguide length is increased. Our GHEM predicts a spectral broadening, but with a different slope (sublinear, e.g., logarithmic) or at a higher intensity level. 

\begin{figure}[h]
    \centering
    \includegraphics[width=85mm]{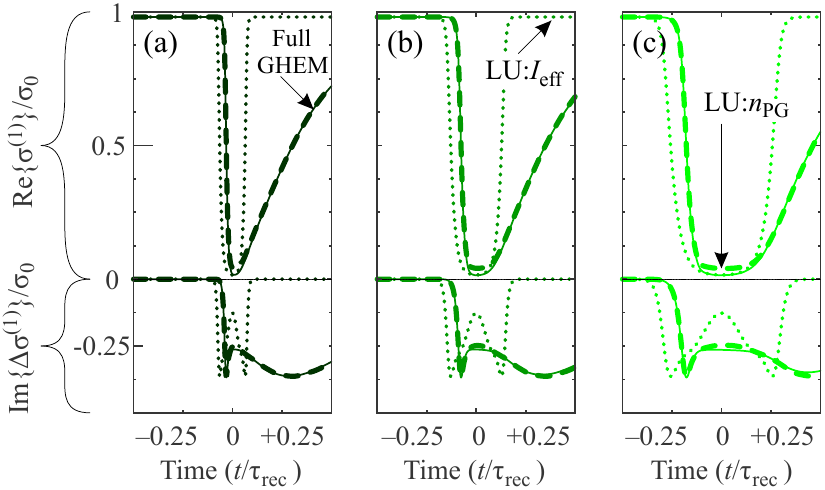}
    \caption{
    {Comparison of full transient photoconductivity response against two `look-up' (LU) strategies based on static photoconductivity. Panels (a,b,c) correspond to pulse duration $\Delta t_\mathrm{FWHM}=(0.5,1,2)$~ps. Thin solid curves are for the full transient GHEM; thick dotted curves are for look-up using directly the pulse intensity; thick dashed curves are for look-up using the GHEM-computed photogenerated carrier density.}
    }
    \label{fig:GHEM_Transient_Compare}    
\end{figure}

Finally, our GHEM shows that both real and imaginary parts of transient $\Delta\sig{1}(t)$ correlate mostly with the $n_\mathrm{PG}(t)$. This means that the carrier density is what predominantly defines the nonlinear response, and validates the phenomenological models proposed in \cite{Vermeulen2018,Demongodin2019}, that only track the carrier-density, and not the carrier temperature and chemical potentials. Nevertheless, we stress that using only the interband rate equation, e.g., Eq.~\eqref{eq:PDE_SAM_nPG}, coupled to the pulse or in static regime, produces results that do not match the full six-equation GHEM response nor the experimental trends. {In Fig.~\ref{fig:GHEM_Transient_Compare} we compare the full transient GHEM response, for the three pulse-widths of Fig.~\ref{fig:GHEM_Transient_VarAll}(f), with a `look-up' (LU) approach using the static saturation curves, e.g., Fig.~\ref{fig:GHEM_CW_varMu}. The dotted and dashed curves in Fig.~\ref{fig:GHEM_Transient_Compare} is the photoconductivity when the look-up is based on the pulse's $\Ieff(t)$ or the resulting $n_\mathrm{PG}(t)$, respectively. The dotted curves fail to capture the edges of the pulse (particularly the trailing edge), while the dashed curves attain a near-perfect fit, with slight difference only on pulse-peak.}

\subsection{Outlook} \label{sec:Outlook}

The quantitative divergences between the model and experiments discussed in the previous subsection hints that more physics should complement the GHEM. In this way, its validity range can be extended to higher effective intensities, such as the ones delivered by high-power fiber lasers into highly-confining waveguides. Photothermal effects, i.e., lattice heating and heat diffusion, provide a preliminary correction to abnormal induced absorption observed at very high intensities, particularly in the carrier saturation regime. Thus, the lattice temperature, $T_0$, can be a problem variable complemented in the GHEM by a differential equation related to Joule heat and its diffusion. Additionally, we theorize that diffusion of carriers \cite{Chatzidimitriou2020}, with a possible dependence of the diffusion coefficient on the carrier temperature according to the Einstein relation, $D_\mathrm{diff}=\mu_\mathrm{mob} k_B T$, could also improve agreement with experimental observations.

This model can be directly used in the MIR/THz spectral region, where the intraband Drude-like contribution dominates the photoconductivity via thermal effects. This regime can be readily implemented within this GHEM, by forcing an appropriately low carrier-density saturation.

Finally, spectral bandwidth studies and/or multi-channel (e.g., pump-probe) effects can be readily incorporated in this GHEM, with the Kubo formulas accounting for the spectra of the model parameters.

\section{Conclusive Remarks} \label{sec:6:Summary}
In summary, we have developed an electrodynamic model for the study of the ultrafast absorptive and refractive transient response of graphene monolayers placed along integrated photonic waveguides. This strongly non-perturbative model agrees with quasi-classical perturbative derivations of third-order nonlinearity in the area of interest, i.e., near and slightly below the half-photon energy, but only at low intensities. For higher intensities, the qualitative agreement with experimental observations, in both static and transient regimes, is good and the model's phenomenological parameters can be fit to measurements for quantitative analysis and design. Apart from the illumination parameters, such as the pulse duration and peak intensity, emphasis was given to the response sensitivity on voltage-tunability (via the equilibrium chemical potential), on the sample's quality (via the impurity density), and on the existence of a carrier-density saturation regime. For most photonic applications, the optimal chemical potential would be close to half-photon energy, while the actual compromise between the coupled absorptive and refractive nonlinearity shift the value slightly above or below it.

\appendix 
\section{Fermi-Dirac Framework}\label{app:FD_Theory}

The graphene hot-electron model (GHEM) adopted from \cite{Mikhailov2019HEM} relies on distinct quasi-Fermi levels (or chemical potentials) for the electron and hole plasmas, $\mu_e$ and $\mu_h$, respectively. In general it holds that $\mu_e\neq\mu_h$, and the difference becomes more pronounced as the system is pushed farther from thermal equilibrium especially by photogeneration (interband absorption). Both plasmas have the same carrier temperature, $T$, which can (far) surpass the lattice temperature $T_0$, if even for a sub-ps timespan. In the Fermi-Dirac statistical framework, the distribution function for electrons/holes is given by
\begin{equation} 
    f_{e/h}(\Ec;\mu_{e/h},T)=\frac{H(\pm\Ec)}{1+\exp\left(\pm\frac{\Ec-\mu_{e/h}}{k_BT}\right)},
    \label{eq:feh}
\end{equation}
where $k_B$ is the Boltzmann constant and $H(x)$ is the step function ($H=1$ for $x>0$, else $H=0$). The values $f_{e/h}(\Ec)$ are color-coded along the vertical (energy) direction of the Dirac cones presented in Fig.~\ref{fig:2:cones}, where the conduction/valence band corresponds to $\Ec\gtrless0$ for this zero-bandgap semiconductor.

The solid-state parameters on which the GHEM finally relies are the carrier surface densities ($n_{e,h}$ in 1/m$^2$) and the plasma-energy surface densities ($\Epd_{e,h}$ in J/m$^2$). Note that, like the distribution functions, these are distinct for electrons ($e$-subscript) and holes ($h$-subscript) and depend solely on the corresponding chemical potential ($\mu_{e,h}$) and the common carrier temperature. The formulas for the carrier and energy densities are derived from energy integrals of $f_{e/h}(\Ec)$ times the density of states which, for graphene near the tip of the Dirac cone, takes the linear form
\begin{equation} 
    N_\mathrm{DOS}(\Ec) = \frac{ 2 }{ \pi (v_F\hbar)^2 }|\Ec|,
    \label{eq:DOS}
\end{equation}
where $\hbar$ is the reduced Plank constant ($\hbar=h/2\pi$) and $v_F \approx c_0/300$ is the Fermi velocity in graphene ($c_0$ is the speed of light in vacuum); formally, $v_F=\alpha_0\gamma_0\sqrt{3}/2\hbar\approx0.874\times10^6$~m/s, for a lattice constant of $a_0=2.46$~\AA~and a nearest-neighbour coupling energy of $\gamma_0=2.7$~eV \cite{Cheng2014,CastroNeto2009}. The quantities $\{n,\Epd\}$ are given by the following expressions: $n =\int_0^\infty\DOS f d\Ec $ and  $\Epd =\int_0^\infty\DOS f \Ec d\Ec$, resulting in: 
\begin{equation} 
    n_{e/h}(\mu_{e/h},T)= \frac {2(k_BT)^2}{\pi(\hbar v_F)^2}F_1\left(\pm\frac{\mu_{e/h}}{k_BT}\right),
    \label{eq:neh}
\end{equation}
\begin{equation} 
    \Epd_{e/h}(\mu_{e/h},T)=\frac {2(k_BT)^3}{\pi(\hbar v_F)^2}F_2\left(\pm\frac{\mu_{e/h}}{k_BT}\right).
    \label{eq:Edeh}
\end{equation}
Note the symmetry in these formulas for electrons and holes, with an attention to the sign used for electrons ($+$) and holes ($-$). The function $F_m(x)$ is the Fermi-Dirac integral (FDI) of order-$m$ defined as
\begin{equation} 
    F_m(x)=\int_0^\infty  \frac {u^m}{1+\exp\left(u-x\right)} du.
    \label{eq:FDI}
\end{equation}
The total carrier and energy densities in a given state (e.g., at thermal equilibrium or quasi-equilibrium) are given by the sum of the electron and hole terms, $n_T=n_e+n_h$ and $\Epd_T=\Epd_e+\Epd_h$, respectively. 

Finally, we stress that Eq.~\eqref{eq:neh} and \eqref{eq:Edeh} are (numerically) invertible, which means that knowing any two variables in the $\{\mu,T,n\}$ or $\{\mu,T,\Epd\}$ set, we can calculate the third variable. In the GHEM developed in this work, we specifically require the inversion of Eq.~\eqref{eq:neh}, for the calculation of the chemical potential when the temperature and carrier density are known,
\begin{equation} 
    \mu_{e/h} = \pm (k_B T) F^{-1}_1\left[ \frac{\pi(\hbar v_F)^2}{2(k_B T)^2} n_{e/h} \right],    
    \label{eq:mu_from_n_T}
\end{equation}
where $F^{-1}_1$ is the inverse function of Eq.~\eqref{eq:FDI} for $m=1$. Note that asymptotic expressions can simplify the numerical inversion in extreme cases, namely $F_1(x)=x^2/2$ and $F_1(x)=e^x$, when $x\gg1$ and $x\rightarrow0$, respectively. Finally, fittings exist for the inverted calculation, i.e., extracting $\{\mu,T\}$ from $\{n_e,n_h\}$, such as the one proposed in \cite{CastellLurbe2020} and its supporting information, applicable in the equilibrium case, $\mu_e\equiv \mu_h=\mu$.

\subsection*{ Temperature Dependence of Plasma Energy Density } \label{sec:Proof_EnergyTemp}

The derivative of the FDI in Eq.~\ref{eq:FDI} is defined as
\begin{equation} 
    \frac{dF_m(x)}{dx}=\frac{\Gamma_F(m+1)}{\Gamma_F(m)} F_{m-1}(x),
    \label{eq:FDIderiv}
\end{equation}
where $\Gamma_F$ is the Gamma function, with $\Gamma_F(2)=1$ and $\Gamma_F(3)=2$ for the cases of interest. This FDI property is useful in evaluating the dependence of the total plasma energy density on carrier temperature. Defining $b=2/(\pi\hbar^2v_F^2)$ and $x_{e,h}=\mu_{e,h}/k_BT$ for shorthand, and using Eq.~\eqref{eq:neh} and \eqref{eq:Edeh}, we have:
\begin{equation} 
\begin{aligned}
    \frac{\partial\Epd_T}{\partial T} &= \frac{\partial}{\partial T}\Big\{ b(k_BT)^3 \left[F_2(x_e)+F_2(-x_h) \right]  \Big\} \\
    &= \frac{3}{T}\Epd_T + \frac{2b(k_BT)^3}{k_BT^2} \Big[-\mu_e F_1(x_e) + \mu_h F_1(-x_h)\Big]\\
    &= \frac{1}{T}\left( 3\Epd_T - \mu_e n_e + \mu_h n_h \right ).
    \label{eq:dEnergy_dTemp}    
\end{aligned}    
\end{equation}

\section{Surface Conductivity Calculation}\label{app:sigmaCalc}

For finite (nonzero) carrier temperatures, the formulas for the intraband ($i$-subscript) and interband ($e$-subscript) surface conductivities entail integration over the energy spectrum \cite{Falkovsky2007}. The full expressions are given below, for the quasi-equilibrium case where the quasi-Fermi levels for electrons and holes can be unequal \cite{Mikhailov2019HEM}, $\mu_e\neq\mu_h$, and assuming arbitrary energy-dependent scattering rates $\Gamma_{i,e}=\Gamma_{i,e}(\Ec)$, as follows
\begin{equation}
\begin{aligned}
    \sig{1}_i&(\omega,\mu_e,\mu_h,T) = \sigma_0
    \frac {i}{\pi  k_B T}
    \int_0^\infty  \frac {\Ec }{\hbar\omega+i\Gamma_i(\Ec)} \\
    &\times\bigg[ {\cosh^{-2}\left(\frac{\Ec-\mu_e}{2 k_B T}\right)} + 
           {\cosh^{-2}\left(\frac{\Ec+\mu_h}{2 k_B T}\right)} \bigg] d\Ec,
    \label{eq:s1i_full}
\end{aligned}
\end{equation}
and
\begin{equation}
\begin{aligned}
    \sig{1}_e&(\omega,\mu_e,\mu_h,T) = \sigma_0
    \frac {4i}{\pi}
    \int_{0}^\infty \frac{\hbar\omega+i\Gamma_e(\Ec)} { [\hbar\omega+i\Gamma_e(\Ec)]^2 - 4\Ec^2} \\
    &\times \left[\frac 1{1+\exp\left(\frac{-\Ec-\mu_h}{k_BT}\right)} -\frac1{1+\exp\left(\frac{\Ec-\mu_e}{k_BT}\right)}\right] d\Ec.
    \label{eq:s1e_full}
\end{aligned}
\end{equation}
In these expressions, the $e^{-i\omega t}$ phase-convention is used, $\sigma_0=q^2/4\hbar\approx61$~$\mu$S is the universal optical conductivity of graphene, and $q$ is the (absolute) electron charge. Note that $\sigma_0$ is responsible for the $2.3$~\% absorption through an air-suspended pristine graphene monolayer, as $0.023\approx1-4/(2+\sigma_0Z_0 )^2$, where $Z_0\approx377~\Omega$ is the free space impedance, according to Eq.~\eqref{eq:TinkhamAbsorption}. For specific information regarding the calculation of the integrals in the $i$- and $e$-conductivities, refer to the dedicated subsections, below.

The integrals in Eqs.~\eqref{eq:s1i_full} and \eqref{eq:s1e_full} can be cast in simple closed-form expressions in the zero temperature limit. However, at room temperatures and above (as is the case in this non-perturbative model), the hyperbolic and exponential functions in square brackets in the integrands, will spread across wider energy bands and thus demand for more careful considerations, even for numerical integration.

\subsection{Intraband Conductivity}\label{sec:s1intra}

The integral in Eq.~\eqref{eq:s1i_full} must be numerically computed in the case where $\Gamma_i$ is strongly energy dependent and/or at high carrier temperatures arising when graphene is absorptive ($\mu_c\ll\hbar\omega/2$) and illuminated with intense optical radiation. However, the integral can be analytically computed when $\Gamma_i$ is assumed constant (energy independent), which is a valid assumption at low carrier temperatures. In this case, the energy spreading of the Fermi-Dirac-like function in the integrand of Eq.~\eqref{eq:s1i_full} is low, i.e., only the part around $\Ec\approx|\mu_c|$ contributes to the final value. So, $\Gamma_i\approx\Gamma_i(|\mu_c|)$ can be assumed constant and the resulting closed-form expression is
\begin{equation}    
    \sig{1}_i = \sigma_0 \frac {4i}{\pi } \frac {k_B T}{(\hbar\omega + i \Gamma_i)}
    \ln\left[\left(1+e^\frac{\mu_e}{k_BT}\right)\left(1+e^\frac{-\mu_h}{k_BT}\right)\right].
    \label{eq:s1i_eh_constGamma}
\end{equation}
Moreover, in thermal equilibrium ($\mu_e\equiv\mu_h=\mu_c$), this expression can be further reduced to the more plainly-evident Drude-like form \cite{Chatzidimitriou2015}
\begin{equation}    
    \sig{1}_i = \sigma_0 \frac {4i}{\pi } \frac {\mu_c}{(\hbar\omega + i \Gamma_i)}
    \mathcal{T}\left(\frac{\mu_c}{2k_B T}\right),
    \label{eq:s1i_constGamma}
\end{equation}
where the temperature-dependent function $\mathcal{T}(x)=x^{-1}\ln[2\cosh(x)]$ tends to $\pm1$ for $|x|>1$, i.e., for $|\mu_c|>2k_BT\approx 52$~meV for room temperature. From Eq.~\eqref{eq:s1i_constGamma} and for the wavelength of interest, $\hbar\omega\approx0.8$~eV (NIR), it can be readily seen that both real (absorptive) and imaginary (refractive) parts of $\sig{1}_i$ are proportional to $|\mu_c|$ while only the real part is significantly affected by $\Gamma_i$, typically smaller than 0.1~eV ($\tau_i>6$~fs): As rate $\Gamma_i$ increases ($\tau_i$ drops) the carriers' momentum relaxation is more efficient thus graphene is less conductive (absorptive), i.e., $\Real{\sig{1}_i}$ diminishes.

\begin{figure}[tb]
    \centering
    \includegraphics[width=75mm]{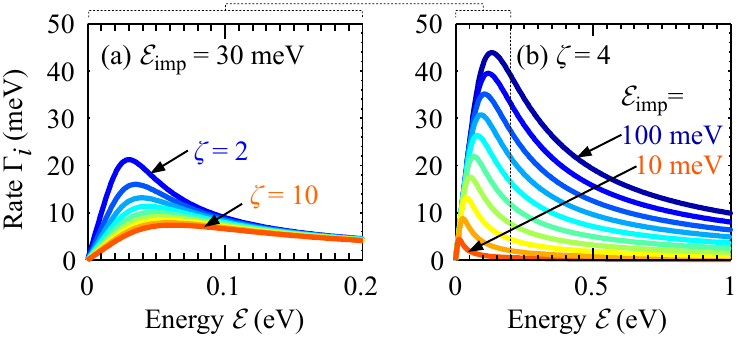}
    \caption{Energy dependence of the intraband momentum-relaxation rate $\Gamma_i$, (a) for various minimal static conductivities $\zeta=2$ to 10 and fixed Coulomb impurity energy $\Eimp=30$~meV, and (b) for various $\Eimp=10$ to 100~meV and fixed $\zeta=4$. Relaxation lifetime: $\tau_\mathrm{[fs]}\approx658/\Gamma_{\mathrm{[meV]}}$.}
    \label{fig:A:Gintra}    
\end{figure}

In this work, we adopt the closed-form model proposed by S.A. Mikhailov in \cite{Mikhailov2019HEM} (and Appendix therein) for the energy-dependence of the intraband scattering rate, which captures the behaviour both at low energies and at high energies, where charged-impurity scattering dominates \cite{DasSarma2011,Trushin2007,Hwang2009}. The proposed formula is
\begin{equation}
    \Gamma_i(\Ec) = \frac{ |\Ec| }{ \zeta/2 -1 + \sqrt{1 + (\Ec/\Eimp)^4} }
    \label{eq:Gintra}
\end{equation}
where $\zeta>2$ is the minimal static conductivity of graphene (at the Dirac point, i.e., when $\{T,\omega,\mu_c\}\rightarrow0$) in units of $\sigma_0$ and $\Eimp$ is a Coulomb energy associated with the density of impurities in the lattice. $\Eimp$ scales with the square-root of the impurity density (typically in the $0.1$-$10\times10^{16}/$m$^{2}$ range) with a proportionality factor depending on the dielectric properties of the medium in contact with graphene; in the same framework it is shown that the low frequency mobility of graphene carriers is inversely proportional to $\Eimp^2$. In any case, both model parameters, $\zeta$ and $\Eimp$, can be fitted to measurements of static conductivity vs. electrically tuned carrier density at low temperatures, or extrapolated from room-temperature measurements. The energy dependence of rate $\Gamma_i$ on $\zeta$ and $\Eimp$ is depicted in Fig.~\ref{fig:A:Gintra}, where the comparatively stronger effect of $\Eimp$ is revealed together with the inverse trend between low energies ($\Gamma_i\propto\Ec$) and high energies ($\Gamma_i\propto\Ec^{-1}$). Note that the quality of a graphene sample increases as its mobility increases, which is inversely proportional to $\Eimp$. In this work, unless otherwise specified, we assume typical values $\zeta=4$ and $\Eimp=30$~meV, corresponding to good quality graphene, with a mobility of about 7000~cm$^2$/(Vs) when the graphene monolayer lies on the interface between air and a $\kappa\approx4$ dielectric substrate.

\subsection{Interband Conductivity}

Unlike the intraband term, the integral in Eq.~\eqref{eq:s1e_full} cannot be analytically solved, even though in most practical cases the interband momentum relaxation rate is energy independent and it can even be neglected, $\Gamma_e\rightarrow0$, under the condition $\Gamma_e\ll k_BT_0\approx26$~meV (at room temperature); in most practical cases $\Gamma_e<1$~meV ($\tau_e>1$~ps) \cite{Gu2012}. This complication is due to the singularity exhibited by the integrand at half-photon energy, $\Ec=\hbar\omega/2$, which can fortunately be circumvented by a transformation involving a principal value integral. This procedure has been outlined in \cite{Falkovsky2008} and, for the out-of-equilibrium case studied here, $\mu_e\neq\mu_h$, it can be extended as follows: We first define the auxiliary function
\begin{equation}
    G(\Ec;\mu_e,\mu_h,T)= \frac{ \sinh\left(\frac{2\Ec-\mu_e+\mu_h}{2k_BT}\right)}
    {\cosh\left(\frac{\mu_h+\mu_e}{2k_BT}\right)+ \cosh\left(\frac{2\Ec-\mu_e+\mu_h}{2k_BT}\right) }
\end{equation}
which can be used in compactly rewriting Eq.~\eqref{eq:s1e_full} as
\begin{equation}
    \sig{1}_e= \sigma_0\frac {4i\Omega}{\pi} \int_{0}^\infty \frac{G(\Ec)}{ \Omega^2-4\Ec^2 } d\Ec,
    \label{eq:s1e_altG}
\end{equation}
where $\Omega=\hbar\omega+i\Gamma_e$ (the scattering rate is assumed energy independent) and the singularity at $\Omega=2\Ec$ is evident in the denominator. Adding and subtracting the term $G(\Omega/2)$ in the nominator of the integrand in Eq.~\eqref{eq:s1e_altG}, we get one singularity-free term [from the $-G(\Omega/2)+G(\Ec)$ terms], that can be straightforwardly numerically computed, and one term that requires a principal-value integral (PVI) [from the $+G(\Omega/2)$ term]. Now, as the integrand function [proportional to $1/(\Omega^2-4\Ec^2)$] is anti-symmetric around the singularity, the PVI reduces to the proportionality constant times $i\pi$. The resulting expression for the numerically integrable interband conductivity is:
\begin{equation}
    \sig{1}_e= \sigma_0\left[  G(\Omega/2) + \frac {4i}{\pi} \int_{0}^\infty \frac{G(\Ec)-G(\Omega/2)}{ \Omega^2-4\Ec^2 } d\Ec\right].
    \label{eq:s1e_integrable}
\end{equation}

Assuming $\Gamma_e=0$, Eq.~\eqref{eq:s1e_integrable} reveals that the real and imaginary parts of the interband conductivity are solely defined by the first and second terms, respectively. Note that the real part can acquire negative values (`gain') owing to population inversion in strongly non-equilibrium states \cite{Mikhailov2019HEM}, i.e., when $\Delta\mu_{(e-h)}>\hbar\omega$; nevertheless, we restrict our study the cases where this regime is not entered, ensuring always that the real part of the total conductivity is positive.

\section{Numerical Solution of Equations System}\label{app:Numerical_Tips}

The GHEM developed relies on a set of differential/algebraic equations (DAE). In its most complicated transient form, the DAE systems consists of two differential equations, \eqref{eq:PDE_AP_Temp} and \eqref{eq:PDE_AP_nPG_Gsat} or \eqref{eq:PDE_SAM_Energy} and \eqref{eq:PDE_SAM_nPG}, and a set of four electroneutrality Eqs.~\eqref{eq:mus_Electroneutrality}. The first set of rate equations has $\{T,n_\mathrm{PG},\mu_e,\mu_h,\mu_{e_0},\mu_{h_0}\}$ as unknowns or variables, whereas the second set has $\{\Epd_T,n_\mathrm{PG},\mu_e,\mu_h,\mu_{e_0},\mu_{h_0}\}$, just as in \cite{Mikhailov2019HEM}. 

The number of unknown chemical potentials in the GHEM can be reduced in a number of ways, (i) by assuming a single Fermi level for electrons and holes in both the hot (out-of-equilibrium) and the QE state, i.e., $\mu_h=\mu_e$ and $\mu_{h_0}=\mu_{e_0}$, or (ii) by assuming the absence of QE state, i.e., $\mu_{e_0}=\mu_e$ and $\mu_{h_0}=\mu_h$, or (iii) by a combination of the previous two cases. This reduces the total equations to four, four, or three, for cases (i)-(iii), respectively, while the carrier temperature (or energy density) and photogenerated carrier density rate equations cannot be dropped, for NIR photons. Even though the qualitative behaviour of the GHEM equation system is equivalent in cases (i)-(iii), we have found quantitative differences in the magnitude of the nonlinear effects and in the intensity threshold between the various regimes. Throughout this work we retain the full six-equation system.

Now, when time derivatives can be dropped, $\partial/\partial t\equiv 0$, we have the static (steady-state) or CW (continuous wave) case, which can be easily handled by MATLAB's FSOLVE. In the more general transient case, e.g., when graphene is excited by a ps-pulse, the DAE can be solved with MATLAB's ODE15S, with an appropriate mass-matrix definition. 

Apart from the six unknowns (variables) and possible time-dependence, one must also define the input parameters (constants) of the system. The most important is the peak effective illuminating intensity, $I_\mathrm{eff}$ in W/m$^2$. Apart from that, and the operating frequency, there is the various material-parameters for graphene: chemical potential ($\mu_c$), lattice temperature ($T_0$), phenomenological intra- and interband lifetimes ($\tau_\epsilon$ and $\tau_\mathrm{rec}$), sample quality (e.g., impurity `energy' $\Eimp$ and residual conductivity $\zeta$), etc.

Finally, all variables and equations should be normalized so that they take near-unity values, to aid numerical solution, either in static or transient mode. The four electroneutrality equations, and the corresponding chemical potentials ($\mu$) are normalized from Joules to electron-volts (eV). The rate equations are normalized to unitless in static (CW) mode and in 1/s in transient (pulsed) mode; in both cases, it is more numerically convenient to compute a normalized temperature $T/T_0$ and photogenerated carrier density $n_\mathrm{PG}/n_{T_0}$.

Refer to the Supplemental Material for MATLAB code implementing the GHEM equation system and its static and transient solution.

\section{Pulse Propagation Along Nonlinear Waveguide}\label{app:NLSE_SSFM}

The initial pulse $B_\mathrm{in}(t)=B(0,t)$ is $z$-stepped through a segment of the waveguide that is sufficiently short ($dz\rightarrow 0$) so that it perturbatively alters the complex amplitude of the pulse envelope; this $z$-stepping is essentially a transient solution of the GHEM, which produces a photoconductivity $\Delta\sig{1}(z,t)$, assumed constant within $[z,z+dz]$, which afterwards distorts the envelope at the end of the step; the procedure is iteratively repeated until the end of the waveguide is reached, applying linear effects (absorption and dispersion) along the way, until we calculate $B(L_z,t)$. This reflectionless step-wise propagation is the essence of the split-step Fourier method (SSFM) \cite{agrawal2012NLFO}. Note that dispersion and linear absorption depend on constant parameters derived from the linear regime eigenmode analysis at $\omega_0$, including the effect of $\sig{1}_\mathrm{lin}$ (primarily on absorption). In contrast, photoconductivity $\Delta\sig{1}(z,t)$ is a nonlinear term, i.e., with temporal dependence, which is moreover updated as the the propagation evolves along $z$, since it is coupled to the pulse envelope, $B(z,t)$. 

The NLSE modeling of linear and third-order nonlinear effects stemming from graphene is well understood \cite{Chatzidimitriou2015}. Now, the GHEM photoconductivity is essentially a free-carrier effect that can be modeled as an extra term in the NLSE \cite{Lin2007,Pitilakis2013}, $\delta_\mathrm{GNL}(z,t)$, together with the dispersion and third-order nonlinearity, whose complex amplitude depends on the spatiotemporal pulse amplitude $B(z,t)$, in a strongly nonlinear manner. A basic single-channel NLSE for the optical pulse envelope $B(z,t)$ (in units of $\sqrt{W}$) modulating a carrier frequency $\omega$, under the $e^{-i\omega t}$ phase convention, is
\begin{equation}
    \label{eq:NLSE}
    \frac{\partial B}{\partial z} =  \bigg( -\frac{\alpha}{2} + D + i\gamma_\mathrm{NL} |B|^2 -\delta_\mathrm{GNL}(z,t)\bigg) B(z,t).
\end{equation}
In this compact form, the constants $\alpha$ and $\gamma_\mathrm{NL}$ are the power loss coefficient (real positive) and the complex third-order nonlinear parameter (including Kerr effect and perturbative SA/TPA), respectively; self-focusing refraction corresponds to $\Real{\gamma_\mathrm{NL}}>0$. Note that $\alpha$ includes a contribution from graphene, through $\Real{\sig{1}_\mathrm{lin}}$, but $\gamma_\mathrm{NL}$ comes exclusively from bulk/3D nonlinear materials and is totally unrelated to graphene. $D$ is the linear dispersion operator, for group-velocity dispersion and higher \{e.g., check Eq.~(4) in \cite{Pitilakis2021}\}, and $t$ is the retarded-envelope time frame, moving with the mode group velocity. Finally, the complex-valued term $\delta_\mathrm{GNL}\propto \Delta\sig{1}$ includes all nonlinear refractive and absorptive contributions from graphene. Notice how $\delta_\mathrm{GNL}$ is added with a minus in Eq.~\eqref{eq:NLSE}, so that an absorption saturation characterized by $\Real{\Delta\sig{1}}<0$ introduces a `gain' that counteracts the linear loss factor $\alpha$. In the same sense, the $\pm$ sign in $\Imag{\Delta\sig{1}}$ corresponds to a defocusing ($+$) or self-focusing ($-$) refraction.

\begin{acknowledgments}
The research work was supported by the Hellenic Foundation for Research and Innovation (H.F.R.I.) under the ``First Call for H.F.R.I. Research Projects to support Faculty members and Researchers and the procurement of high-cost research equipment grant.'' (Project Number: HFRI-FM17-2086)
\end{acknowledgments}

\bibliography{Bib_GNL_2022} 

\end{document}